\documentclass{egpubl}
 
\JournalPaper         
\usepackage[T1]{fontenc}
\usepackage{dfadobe}  
\usepackage{amssymb}
\usepackage{cite}  
\usepackage{subcaption}
\usepackage{soul}
\usepackage{xcolor}

\BibtexOrBiblatex
\electronicVersion
\PrintedOrElectronic
\ifpdf \usepackage[pdftex]{graphicx} \pdfcompresslevel=9
\else \usepackage[dvips]{graphicx} \fi

\newcommand{\revision}[1]{{\color{black} {#1}}}
\newcommand{\deleteforrevision}[1]{}

\usepackage{egweblnk} 

\usepackage{xspace}

\newcommand{\myparagraph}[1] {\mbox{ }\\ \noindent{\textbf{#1}}}
\newcommand{\etal}{et al.\xspace}

\newcommand*{\inlinegraphics}[1]{%
    \raisebox{-.1\baselineskip}{%
        \includegraphics[
        height=0.8\baselineskip,
        keepaspectratio,
        ]{#1}%
    }%
}

\title[A Scalable System for Visual Analysis of Ocean Data]%
      {A Scalable System for Visual Analysis of Ocean Data}

\author[Jain \etal]
    {\parbox{\textwidth}{\centering 
    Toshit Jain$^1$\orcid{0000-0003-0702-3435},
    Upkar Singh$^1$,
    Varun Singh$^1$,
    Vijay Kumar Boda$^1$,
    Ingrid Hotz$^{2,1}$,
    Sathish S. Vadhiyar$^3$,\\
    P. N. Vinayachandran$^4$,
    Vijay Natarajan$^{1,5}$\orcid{0000-0002-7956-1470}
    }
    \\
    \parbox{\textwidth}{\centering 
    $^1$Department of Computer Science and Automation (CSA), Indian Institute of Science Bangalore, India\\
    $^2$Department of Science and Technology (ITN), Link\"oping University, Norrk\"oping, Sweden\\
    $^3$Department of Computational and Data Sciences (CDS), Indian Institute of Science Bangalore, India\\
    $^4$Centre for Atmospheric and Oceanic Sciences (CAOS), Indian Institute of Science Bangalore, India\\
    $^5$Zuse Institute Berlin, Germany\\
    }
    }

\begin{document}


\maketitle
\begin{abstract}
Oceanographers rely on visual analysis to interpret model simulations, identify events and phenomena, and track dynamic ocean processes. The ever increasing resolution and complexity of ocean data due to its dynamic nature and multivariate relationships demands a scalable and adaptable visualization tool for interactive exploration. We introduce pyParaOcean, a scalable and interactive visualization system designed specifically for ocean data analysis. pyParaOcean offers specialized modules for common oceanographic analysis tasks, including eddy identification and salinity movement tracking. These modules seamlessly integrate with ParaView as filters, ensuring a user-friendly and easy-to-use system while leveraging the parallelization capabilities of ParaView and a plethora of inbuilt general-purpose visualization functionalities. The creation of an auxiliary dataset stored as a Cinema database helps address I/O and network bandwidth bottlenecks while supporting the generation of quick overview visualizations. We present a case study on the Bay of Bengal (BoB) to demonstrate the utility of the system and scaling studies to evaluate the efficiency of the system.



\end{abstract}  

\section{Introduction}\label{sec:introduction}
Oceanography refers to the study of the physical, biological, and chemical features and characteristics of the ocean. A comprehensive study of the oceans of the earth has implications for scientific understanding, environmental management, and societal well-being. \revision{It is crucial for predicting extreme events like hurricanes and tsunamis, enhancing our understanding of large-scale planetary processes such as global warming, and ensuring the sustainable management and preservation of ocean resources and marine ecosystems.} With the advancements in collection and generation of ocean data~\cite{rosenblum1989visualizing, fraser2006data}, there is a demand for scalable tools that support effective and interactive visualization of ocean data. Data in oceanography typically consists of multivariate 3D spatiotemporal fields. The fields are generated using simulations or available from satellite imagery, buoy sensors, or in-situ physical observations. Thanks to the explosion in the capability of modern computers and imaging technology, the size of ocean datasets is becoming larger and larger. As a result, visualizing these datasets at interactive speeds is a challenging problem. Further, the datasets are seldom stored locally. The storage on remote servers results in an additional time cost that is dependent on the bandwidth. The dynamic nature of the multiple scalar and vector fields representing physical quantities pose further challenges to the development of efficient visualization methods. 
Among the various fields, ocean currents stand out as an important field of interest. Ocean current is a predominant factor in maintaining the heat equilibrium of the ocean-atmosphere system and in influencing the transport of minerals and salt. 

Several complex structures such as eddies and surface fronts are studied in oceanography. Mesoscale eddies~\cite{robinson2010mesoscale}, circular spiral-like oceanic current pattern that typically span tens to hundreds of kilometers in diameter and last for days to months, are a salient feature of the ocean and vital for ocean analysis\cite{mcwilliams2008nature, mcneil1999new, benitez2007mesoscale}. \revision{A \textit{surface front} is the boundary between two or more volumes of water with distinct temperature or salinity characteristics.} It is represented as a subset of the boundary of a temperature or salinity isovolume. Understanding its movement through 3D space and time is crucial for analyzing ocean dynamics and processes. There also exist some submesoscale features that play a role in the study of the movement of particles within the ocean. Submesoscale features are transient features that span over 0.1 to 10~km and can last between a few hours to a few days. The submesoscale features generate spatial and temporal variations in salinity, temperature, and density that create diverse habitats for different marine organisms, dictating their distribution and abundance in the ocean. Submesoscale currents also help in understanding and predicting the movement and spread of particles in the ocean like nutrients and pollutants. One of the most prominent and useful submesoscale features is called a \emph{filament}, a thin and elongated isovolume of water most often found at the fringes and boundaries of an eddy or along the coast. Since the scale of these features is very small and transient, they are only visible in high-resolution ocean models with a resolution of about \textasciitilde1~km. 

\revision{This paper addresses the need for scalable solutions for 3D ocean visualization. It introduces pyParaOcean, a visual analysis system that} leverages the power of the widely used open source visualization engine ParaView~\cite{AhrensGeveciParaview2005} to enable scalable visualizations of data available from ocean models while supporting a multitude of tasks and functionalities that are specialized for oceanography.

\subsection{Related work}\label{sec:relatedworks}
Visualization in oceanography is a challenging area of research due to the rapidly increasing size, heterogeneity, and multivariate nature of the data, and the inherent complexity of ocean phenomena. The use of general purpose analysis and visualization software such as Matlab, Tecplot, AVS, and ParaView is prevalent in the community. However, oceanographers often use tools developed specifically for ocean data, such as Ferret~\cite{ferret}, pyFerret~\cite{pyferret}, and Copernicus MyOcean~\cite{myocean}. These specialized tools are often developed as standalone solutions and further produce 2D views of the data. 

Some software frameworks developed within the visualization community provide 2D and 3D data visualization capabilities.  COVE~\cite{grochow2008cove} is a collaborative ocean visualization environment that supports interactive analysis of ocean models over the web. RedSeaAtlas~\cite{afzal2019redseaatlas}  supports the selection of regions in a 2D map and provides exploratory views of winds, waves, tides, chlorophyll, etc. over the Red Sea. OceanPaths~\cite{nobreocean2015} is a multivariate data visualization tool that computes pathways tracing ocean currents and supports the plotting of different high-dimensional data along the pathways. This enables the study of correlations between different oceanographic features. \cite{hua2023hybrid} developed a 3D eddy identification technique based on the sea surface height and the velocity field. Sea surface height and velocity profile have also been previously used for eddy detection~\cite{matsuoka2016new}.

An oceanographer's analysis workflow includes a few common tasks~\cite{grochow2008cove} such as inspection of temperature and salinity distributions and vertical cross sections, compare recently measured salinity data against model data, inspect and analyze current vorticities and circulation based on flow data, and analyze extreme events. Isosurfaces and volume rendering are natural choices for visualization of 3D temperature and salinity distributions~\cite{dinesha2012uncertainty,park2004visualization}. However, visualization of dynamically changing distributions is a challenge.  VAPOR~\cite{li2019vapor} is one of the few tools that provides efficient 3D visualization for oceanography and atmospheric science applications. The VAPOR data collection (VDC) model supports interactive visual analysis of large data sizes on modern GPUs and commodity hardware.  

Xie \etal~\cite{xie2019survey} and Afzal \etal~\cite{afzal2019state} present surveys of visual analysis methods and tools developed for ocean data. Xie \etal classify the visual analysis tasks into four categories: study of different environmental variables, ocean phenomena identification and tracking, discovery of patterns and correlations, and visualization of ensembles and uncertainty. Further, they identify opportunities and unexplored areas for future work including efficient and scalable methods for data processing and management, identification of features at multiple scales,  and immersive platforms. 

Singh \etal~\cite{singh2022frontskeleton, singh2024advection} used geometric and topological descriptors to track high salinity water. The study showed that upon entering the Bay of Bengal (BoB), the high salinity water mass splits into three major directions and advances toward Visakhapatnam, the coast of the Andaman and Nicobar Islands, and the center of BoB. In this paper, we report an improved fast and parallel implementation of the salinity front tracking method for large data sizes to demonstrate that the approach is scalable.

\subsection{Contributions}
We present pyParaOcean, a scalable system for visual analysis of ocean data. The visualization capabilities of pyParaOcean are available via a seamless integration into ParaView~\cite{AhrensGeveciParaview2005} using plugins. Further, the server-client architecture of ParaView is leveraged to scale the computations and visualizations to large data sizes. pyParaOcean also offers a Cinema Science database generator~\cite{ahrens2014image} to enable quick analysis and overview generation of a dataset. Using a Cinema database helps to bypass the initial setup time needed for launching a ParaView server on multiple cores and distributing a dataset on those cores, when the goal of the user is to simply to generate an overview. An advantage of this approach is that it offers the flexibility to jump into ParaView for subsequent in-depth analysis and visualization. Main contributions of this paper include
\begin{itemize}
\item \revision{pyParaOcean: A scalable, extensible, and three-dimensional visual analysis system for ocean data that offers support for seeding strategies for fieldline computation, generating parallel coordinate and vertical section views, and for computing and visualizing eddies and fronts.}
\item A parallel implementation of an algorithm for front-based tracking of salinity movement~\cite{singh2022frontskeleton}.
\item A detailed scaling study of the pyParaOcean modules for extracting and visualizing complex mesoscale structures, including eddies and surface fronts.
\item A Cinema Science database generator that enables quick analysis and overview visualization via slices along time and depth.
\item A case study on a high resolution dataset to demonstrate the need for a scalable ocean analysis system.
\end{itemize}
The case studies presented in this paper, performed in collaboration with an oceanographer, focus on the exploration of the salinity distribution within the BoB and demonstrate the utility of the proposed application-centric scalable visualization system.  A preliminary version of this paper introduced pyParaOcean as a visual analysis tool~\cite{jain2023pyparaocean} and outlined its key functionalities and features. This paper  presents a detailed description of the system, describes additional functionalities, new parallel implementations, a detailed scaling study, improvements in I/O via the generation of a Cinema Science database, and a new case study on a large dataset.
\begin{figure*}
    \centering
    \includegraphics[width=\linewidth]{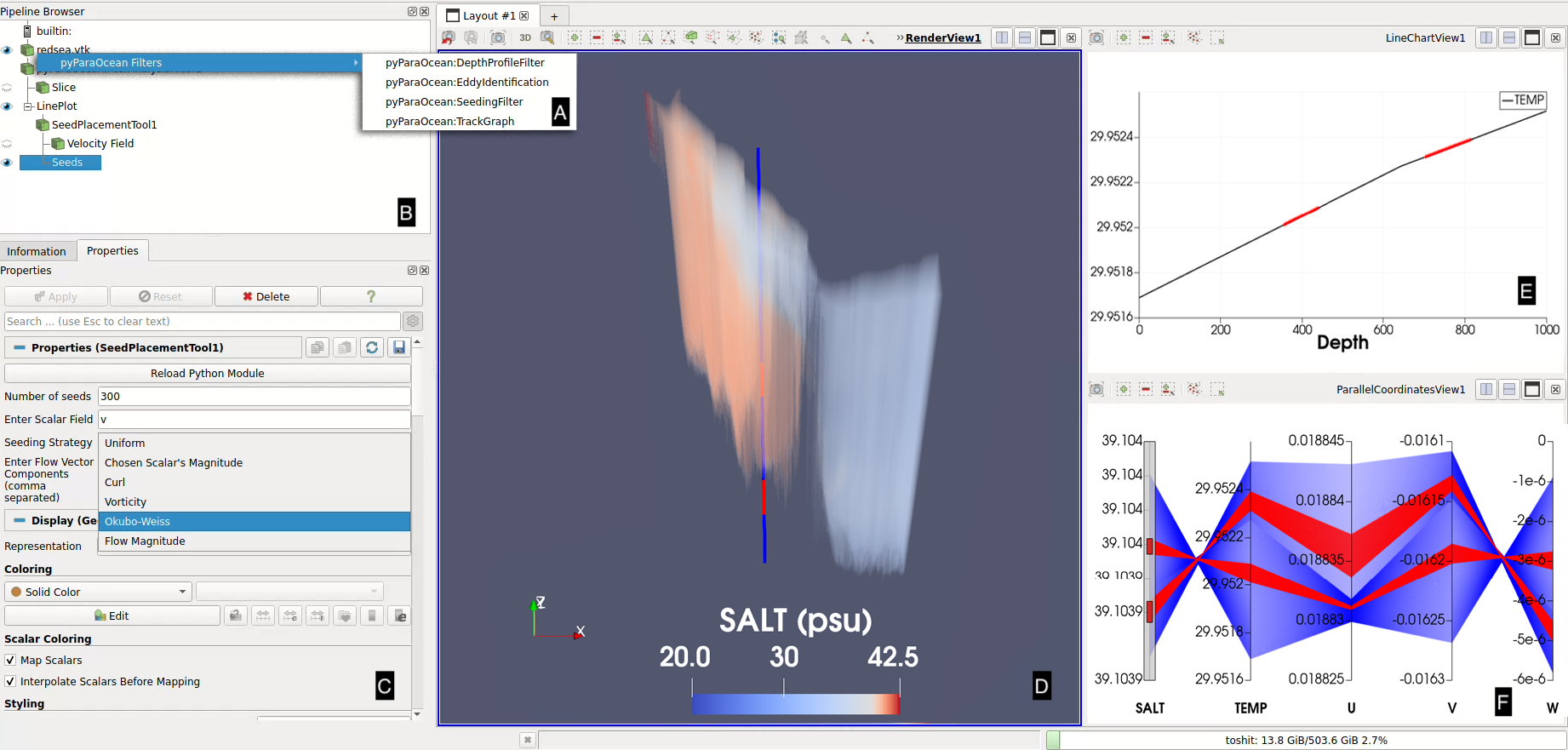}
    \includegraphics[width=.24\linewidth]{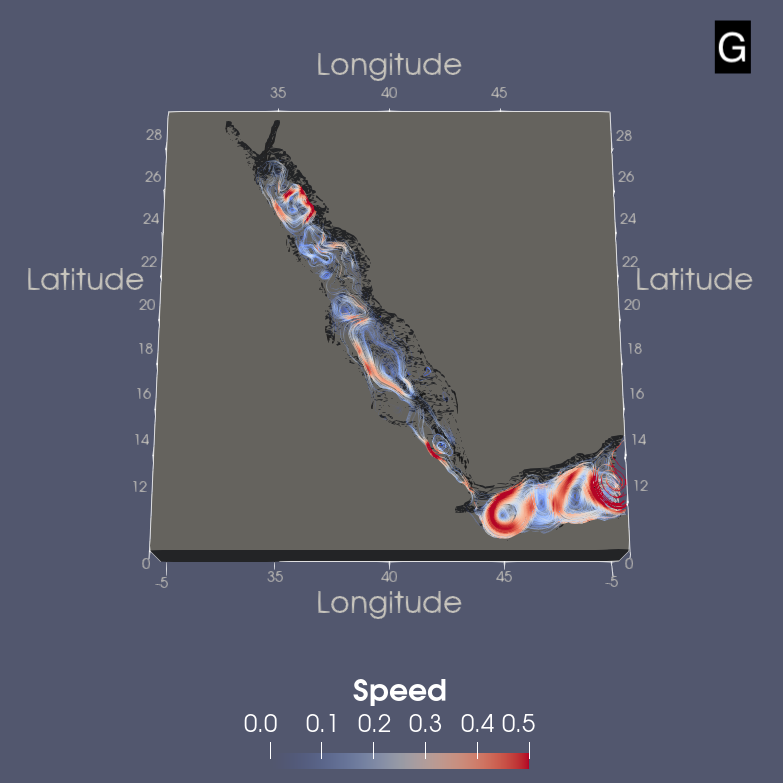}
    \includegraphics[width=.24\linewidth]{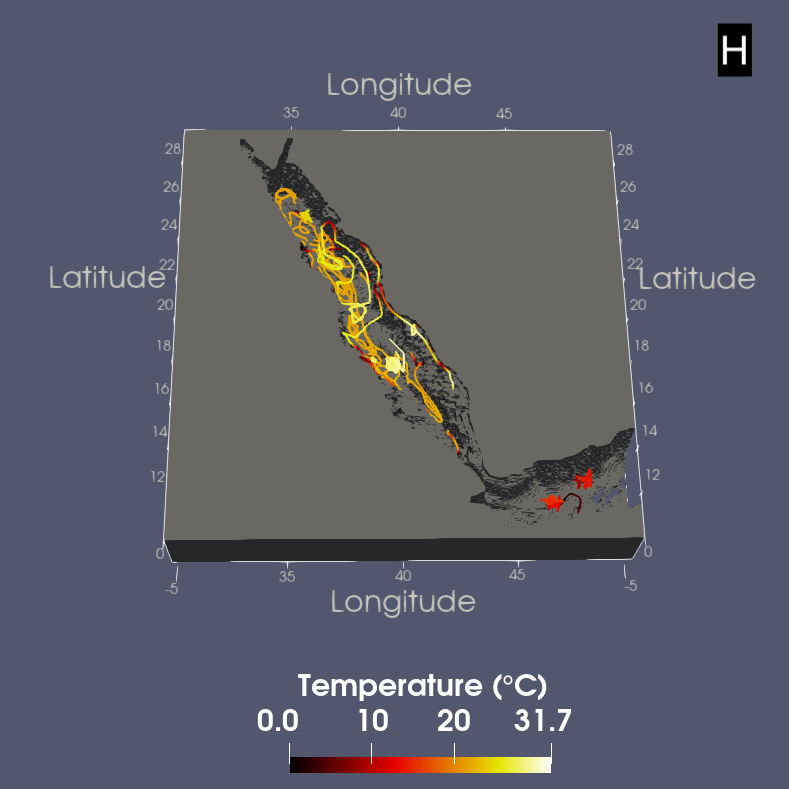}
    \includegraphics[width=.24\linewidth]{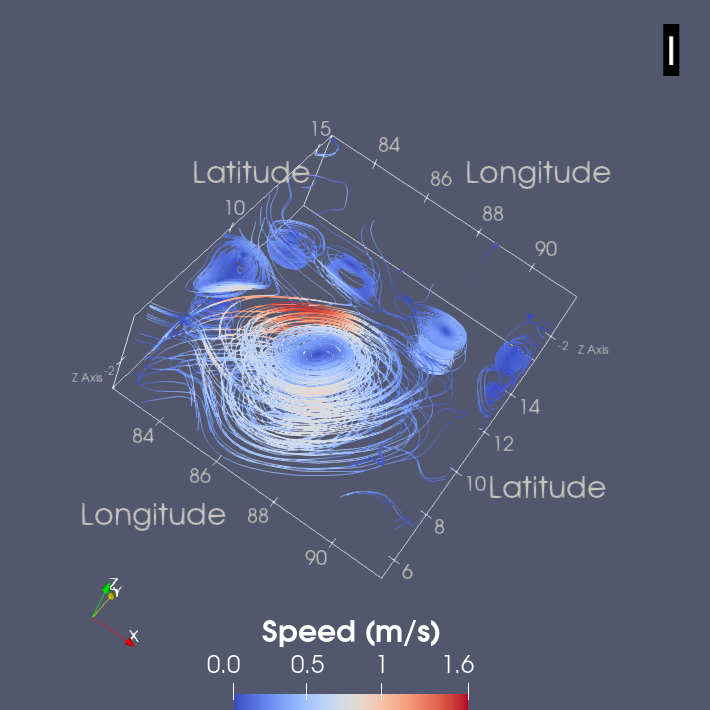}
    \includegraphics[width=.24\linewidth]{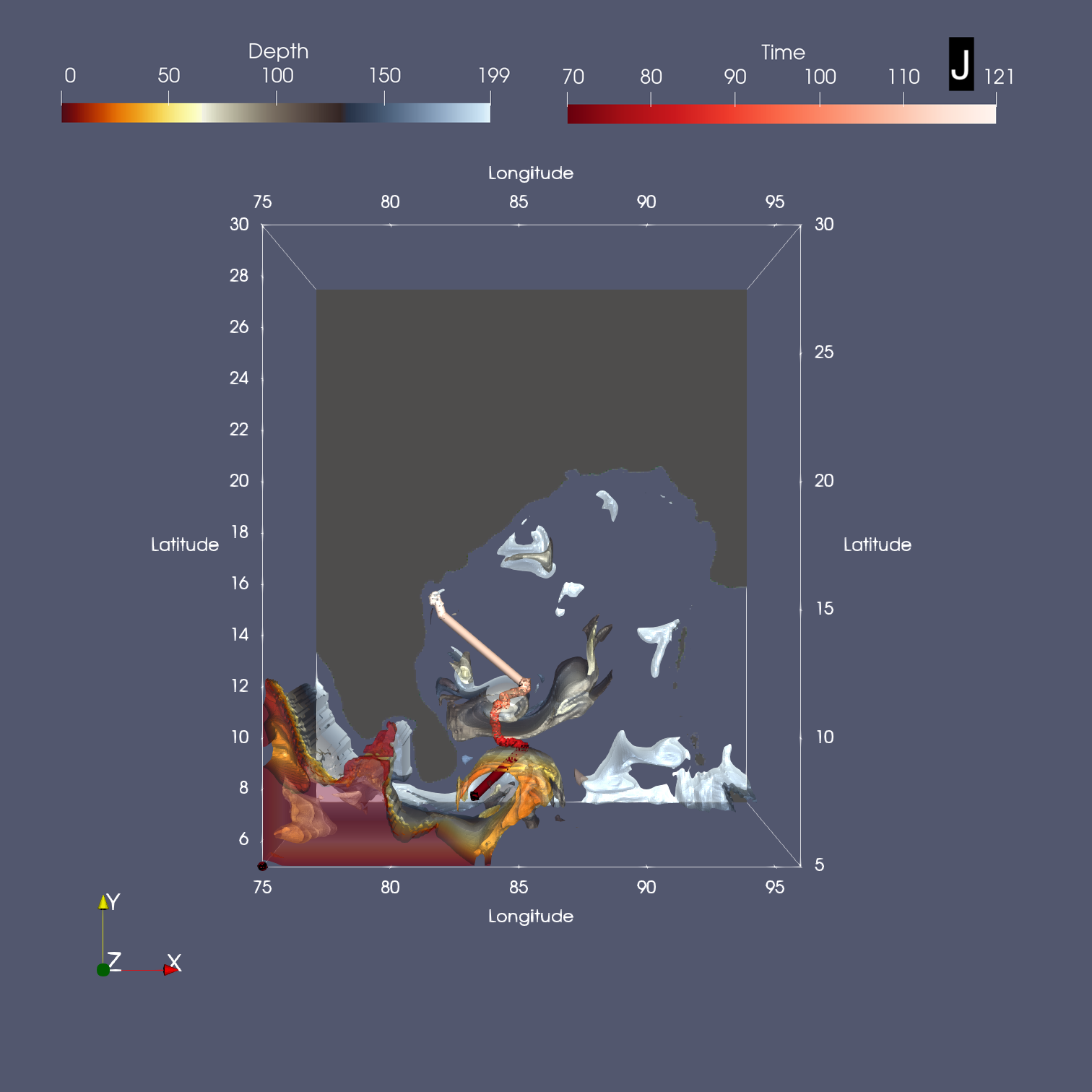}
    \caption{pyParaOcean functionality and user interface. (A)~All pyParaOcean modules are implemented as ParaView filters. (B)~ParaView pipeline browser shows the different datasets under study and the filters applied on them. (C)~The seeding filter from pyParaOcean provides multiple options for tracing fieldlines. The figure illustrates the usage of various filters showcasing (D)~salinity visualization using volume rendering, (E)~interactive depth profile query visualization, (F)~multivariate data visualization using a parallel coordinates plot of all fields in the dataset, (G)~flow visualization with streamlines, (H)~interactively seeded pathline visualization, (I)~eddy detection and visualization, and (J)~tracking high salinity water movement with a surface front track.}
    \label{fig:pyparaocean}
\end{figure*}

\section{Data}\label{sec:data}
Oceanography involves the study of intricate temporal and spatial processes that encompass interactions between entities in various scales. The analysis spans from smaller scale features such as eddies and fronts to large-scale structures such as ocean basins and circulation patterns~\cite{xie2019survey}. Ocean data often consists of a collection of time-varying scalar and vector-valued fields on 3D domains. The data is available from simulations, satellite imagery, buoy sensors, or in-situ observations. The large data sizes are due to the availability of high-performance computing and storage resources, increased sampling resolution, and a growing number of observables. The various fields are available as samples on a rectilinear grid, also called a latitude-longitude gird, in the NetCDF format \cite{netCDF1990interface}. \revision{While the description below is restricted to rectilinear grids, the filters in pyParaOcean extend to data available on other grids that are supported by ParaView.} All visualizations and analysis in the paper is performed on data over the Bay of Bengal region available from two sources, GLORYS12V1 and ROMS. Both are \emph{reanalysis} datasets, which integrate numerical simulation models with observational inputs thereby ensuring spatiotemporal consistency.

\myparagraph{Global Ocean Physics Reanalysis (GLORYS12V1).}
This dataset~\cite{Madec2008} is a reanalysis product and provides multiple fields including salinity and horizontal velocities across latitude and longitude in NetCDF format. The salinity field is considered during the period June 2016 -- September 2016 at daily resolution (122 time steps) on a 3D rectilinear grid, regularly sampled horizontally with a latitude-longitude resolution of $1/12^{\circ}$ and irregularly sampled across depth at 50 levels. The data is clipped to the geographical region corresponding to the BoB, specified as bounds on longitude ($75^{\circ}E$ to $96^{\circ}E$) and latitude ($5^{\circ}S$ to $30^{\circ}N$), using the command line tool Climate Data Operators (CDO)~\cite{schulzweida2019}.  Further, the salinity field is considered up to a depth of 200~m and resampled at regular depth levels.  The resampling computation is scheduled in parallel for each time step. High salinity water movement is observed only in relatively shallow waters~\cite{AnutaliyaSend2017} and hence the restriction. The resulting NetCDF file is used for all further processing and analysis. We refer to this dataset as the GLORYS dataset henceforth. The scalability analysis of the front-based salinity movement tracking algorithm requires data at different resolutions. This is created by resampling the salinity field at regular intervals of depth, latitude, and longitude using linear interpolation. The samples are at depth levels 1~m apart up to 200~m depth, and at a latitude-longitude resolution of $1/(12\times r)^{\circ}$, $r \in \mathbb{Z}$. The resulting 3D regular grid data enables efficient volume processing and visualization. 

\myparagraph{Bay of Bengal ROMS.}
This dataset is generated from the well-established high-resolution Regional Ocean Modeling System (ROMS)~\cite{shchepetkin2005regional}. ROMS is a free surface, three-dimensional primitive equation ocean circulation model based on non-linear sigma (${\sigma}$) coordinate of~\cite{song1994semi} and widely used for a diverse range of regional ocean applications. The model configured for the BoB basin stretches from 77$^{\circ}$E to 99$^{\circ}$E in the zonal direction and from 4$^{\circ}$N to 23$^{\circ}$N in the meridional direction with spatial resolution of 1/96$^{\circ}$ (${\approx}$1km in zonal direction). There are 40 vertical depth (${\sigma}$) levels. The vertical levels are allocated in such a way that the vertical resolution is highest in the top 150~m of the water column. For initial and boundary conditions, the HYbrid Coordinate Ocean Model (HYCOM) 1/12$^{\circ}$ daily reanalyses data are used~\cite{bleck2002oceanic, halliwell2004evaluation, chassignet2007hycom}. The model was initialized on January 1, 2012, and integrated till December 31, 2013, without any salinity relaxation. For presenting the scaling studies, the dataset is interpolated to the same 27 depth levels as the GLORYS dataset to present comparable results. 
Again, the scalability analysis of the front-based salinity movement tracking algorithm requires data at different resolutions. This is created by resampling the salinity field in all 240 time steps (daily resolution beginning June 2012). Again, the samples are at depth levels 1~m apart up to 200~m depth, and at a latitude-longitude resolution of $1/(12\times r)^{\circ}$, $r \in \mathbb{Z}$.

\section{pyParaOcean: Design and architecture}\label{sec:design and architecture}
In this section, we describe the system design and architecture of pyParaOcean, a tool built upon ParaView~\cite{AhrensGeveciParaview2005} and designed to support visualization tasks in oceanography. ParaView is an open-source visualization software built upon VTK~\cite{vtkBook} that enables the creation of a visualization pipeline from a network of executable modules. Each module within ParaView can be considered as a functional unit, featuring various input and output ports. A module can perform one of three functions: data generation (zero input ports; one or more output ports), perform some computation or transformation on the incoming data (multiple input and output ports), or render and produce images or graphic primitives (no output ports). While ParaView is a versatile visualization tool equipped with an extensive array of readers, data sources and filters, the sheer volume of available filters can become overwhelming and challenging to navigate, especially for experts in specific application domains. Additionally, ParaView offers the flexibility to incorporate new modules through python plugins, ensuring that users can adapt the software to their needs when the existing set of modules are inadequate.

\begin{figure}
  \centering
  \includegraphics[width=\linewidth]{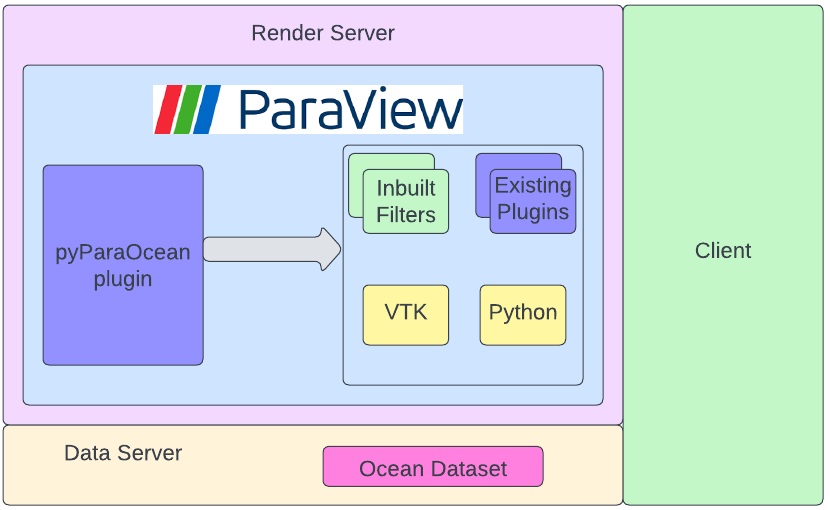}
  \caption{pyParaOcean system architecture.}
  \label{fig:architecture}
\end{figure}

Figure~\ref{fig:architecture} shows the architecture of pyParaOcean, which comprises of data parallel \textit{filters} specifically designed to offer visualization capabilities for the interactive exploration of the three-dimensional ocean data. This design supports easy incorporation of new functionalities as filters in ParaView, the parallel execution of the data server and the render server, and  interactive exploration through the client.

\subsection{Parallelism}
%
ParaView provides support for parallel data processing, which is helpful for handling large and high-resolution datasets. ParaView can be deployed in a server-client manner. The server stores all data, plugins, and is responsible for rendering, reading, writing, and computing. It can be deployed on a cluster or a supercomputer in a parallel manner using OpenMPI. Moreover, the server can be split into a data server for handling all data processing tasks and a render server for handling all rendering tasks. Each server can be deployed on a different compute hardware. Both the data server and the render server are collectively referred to as \textit{pvserver}. All interactions are handled at the client, which drives the process by creating the visualization pipeline and displaying the generated visualizations.

\subsection{Load balancing and ghost cells}
\begin{figure}
  \centering
  \includegraphics[width=\linewidth]{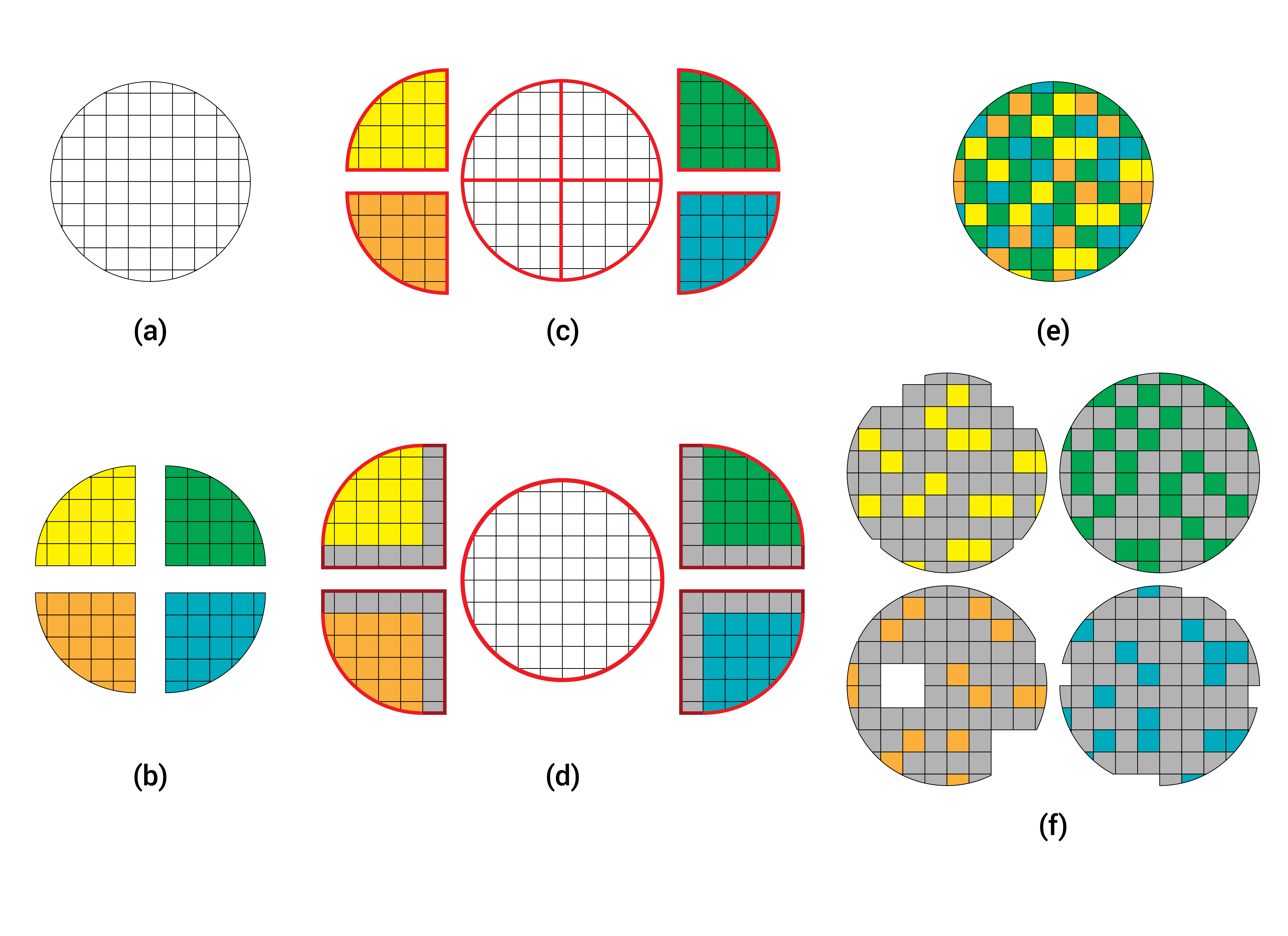}
  \caption{(a)~An example 2D mesh. (b)~The mesh is partitioned for parallel processing. Four processes are created and each chunk is sent to a different process. Each process is represented by a unique color (\inlinegraphics{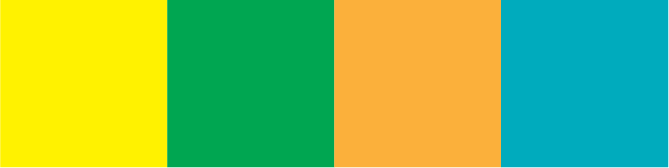}). The \texttt{vtkDataSetSurfaceFilter} filter is applied to the dataset to compute its boundary. (c)~Output of the filter when there are no ghost cells. The filter incorrectly reports edges from the interior of the data as boundary. (d)~Ghost cells are inserted on the common boundary between the individual partitions. The filter now reports the correct output, since all false positives are attached to the ghost cells, which are eliminated from the final output. (e)~An alternate partitioning of the data into four processes. (f)~After the addition of a ghost cell layer, it is apparent that this is an inefficient partitioning of the data since all processes work on almost the entire data, due to poor load balancing.}
  \label{fig:ghostcells}
\end{figure}

ParaView partitions the data into a number of chunks that is equal to the number of parallel processes. Each partition is sent to a different process, and the processes compute the filter on the partition assigned to them independently. VTK and ParaView are designed to keep the communication between processes to a minimum and do not exchange data with each other. This is one of the reasons for the efficiency of ParaView in a distributed setting. Since data is not shared between processes, ParaView uses a concept called \textit{ghost cells} to ensure that the filters produce the correct output. A ghost cell is a data item (say, a voxel in a 3D grid) that belongs to one process but is duplicated on another process corresponding to a spatial neighbor that shares a common boundary. These cells are read-only for the process. While the data is available to the process, the cell is ``owned'' by another process. Figure~\ref{fig:ghostcells} provides an example to illustrate the idea. The partitioning of the data also dictates the load balancing and performance of a filter. Partitioning and load balancing assume high importance in the context of ocean data. If the data contains landmass, a region where no computation takes place, the process that holds the chunk of data with little or no ocean data will run idle and result in an inefficient load balancing. 

\section{pyParaOcean: Functionalities}\label{sec:functionalities}
In this section, we list the functionalities and modules available within pyParaOcean. These modules for ocean data visualization are implemented as plugins and filters in ParaView. Figure~\ref{fig:pyparaocean} shows the output of the modules and the user interface of pyParaOcean. 

\subsection{Seed placement and fieldlines visualization}
Fieldlines, which encompass streamlines and pathlines, offer a comprehensive perspective of a 3D vector field. pyParaOcean provides a filter to choose from diverse seeding strategies for streamline and pathline computation. The seeds produced by this filter serve as input for the customized source streamline integrator or particle tracer within ParaView (Figure~\ref{fig:pyparaocean}(G,H)).
Streamlines are a collection of integral curves aligned with the velocity vector at each spatial point, serving as instantaneous representations of flow patterns that help understand significant oceanographic occurrences like eddies, currents, and filaments. Pathlines trace the evolution of the velocity field over time, by depicting the trajectory of a massless virtual particle from a seed point at a specific time step. Pathlines are helpful in comprehending transport phenomena such as salinity advection and debris accumulation. Generating the pathlines is compute intensive compared to streamlines, and is heavily dependent on I/O speed. 

The seeding filter controls the number of seeds and manner in which the domain is sampled for seed placement, see Figure~\ref{fig:pyparaocean}~(C). Sampling can be achieved through (a)~uniform distribution, (b)~weighted based on flow speed, curl, vorticity, or the Okubo-Weiss criterion, or (c)~weighted according to user defined scalar fields that are calculated earlier in the pipeline. Users have the flexibility to fine tune various line integration parameters and sampling preferences to minimize visual complexity, focus computations on specific areas of interest, maximize domain coverage, and emphasize interesting flow characteristics.

\subsection{Isovolumes and isosurfaces}
Volume rendering is a natural choice for visualizing 3D scalar fields because it provides a quick overview of the distribution. The volume rendering filter in ParaView can be customized to visualize subvolumes of interest by choosing an interval within the range of the scalar field. For instance, a user may extract an isovolume containing the mean salinity / temperature value within the spatial region of interest, or an isovolume that captures high salinity water. 

\subsection{Depth profile view} \label{subsec:depthprofile}
This filter enables the user to inspect a vertical column of the ocean, specified by a longitude and latitude pair.  It drops a ``needle'' into the ocean and samples points along this vertical line at different values of depth. A linked parallel coordinates plot presents a depth profile of all scalar fields sampled along the vertical column. A line plot view of the chosen scalar field against increasing depth values is also displayed. Optionally, the scalar field mapped to a vertical slice at the chosen longitude is shown in the volume render window. The user can select and highlight a subset of points in the vertical column from the parallel coordinates plot and track them across time in all views. This is useful for studying vertical mass transport, especially upwelling or downwelling via Ekman transport~\cite{sarmiento2013ocean} in eddy centers, and to study the depression of isotherms indicating redistribution of heat~\cite{kumar2007eddy}. 

\subsection{Front tracking}
Oceanographers frequently study water masses that are responsible for mass or heat transport. The water masses are often characterized by distinct temperature or salinity ranges. Singh \etal~\cite{singh2022frontskeleton} developed novel representations of high salinity water using connected components of an isovolume boundary, called the \emph{surface front}. The front tracking filter computes the surface front, traces their movement over time, and generates a track graph summarizing the movement of all surface fronts. The filter uses the Python multiprocessing library for parallel processing and can therefore execute only on the cores present on the local workstation (client). In order to support larger data sizes, we have developed a stand-alone Python script that uses MPI to utilize all cores in a cluster to compute the track graph. A selected set of tracks derived from this graph can be displayed for visual analysis. Surface fronts have proven to be effective representations of high salinity water masses and have been utilized to trace the movement of high salinity water as it enters the BoB from the Arabian Sea.

\subsection{Eddy identification and visualization}\label{subsec:eddyidentificationtool}
The eddy identification filter in pyParaOcean is designed to be extensible and allows for multiple implementations.  The current implementation of the filter focuses on mesoscale eddies~\cite{amores2017coherent}. It uses  the velocity field in individual time steps and does not require any derived fields. This 3D detection scheme can be applied in parallel across time steps and across depth slices since the vertical velocity is not used. The flow speed of the swirling fluid decreases radially inward towards the center of rotation. The filter inspects the local minima of the flow speed to identify potential eddy centers. Vertical velocity is ignored to discount the motion of upwelling or downwelling in vortex cores, thus enhancing the corresponding flow minima. Noise and less significant minima are removed by applying a topological persistence-driven simplification~\cite{tierny2017topology}. Next, the method employs an approximation of the winding angle criterion~\cite{friederici2021winding} by checking if the streamline crosses into all four quadrants of an XY plane that is centered at the minimum~\cite{daguo04cross}. This method is more effective in regions with relatively stationary eddy centers. Streamlines seeded near the core of an eddy form spirals or closed loops. The boundary of an eddy is determined using a binary search along the radial axes. The search helps locate the seed that is furthest from the eddy center but results in a spiral or nearly closed loop streamline. The filter displays all streamlines originating near the detected vortex core, and hence presents a 3D profile of the eddy. 

There is potential for implementation of  other existing methods for reliable eddy identification and visualization. For example, a vorticity based method~\cite{mcwilliams1990vortices} that identifies the eddy centers using vorticity extrema and calculates the eddy boundary by comparing the neighborhood vorticity values to the center, or one that uses a special Okubo-Weiss parameter based on shear and strain deformation and the vertical component of vorticity to measure rotation and hence identify potential eddies~\cite{okubo1970horizontal}. The filter may be extended to support other eddy detection methods~\cite{matsuoka2016new,friederici2021winding} that may be selected via the interface.

\subsection{I/O and the NetCDF format}
Ocean data is typically stored in the NetCDF format. NetCDF (Network Common Data Form) is a versatile file format and software library widely used in scientific research, particularly in areas such as atmospheric science, climate science, and specifically within oceanography. It is self-describing, which means that the file contains metadata, such as data attributes, dimensions, and variable specification that describe the structure and content of the data. It offers a hierarchical structure for complex data, is platform-independent, supports large datasets efficiently, and is accessible through various programming languages. Since this data format is self-describing, metadata handling can become an I/O bottleneck for the NetCDF format since the metadata is often stored in a single location within the file system. 
Beginning with version~4, NetCDF supports parallel I/O, built on top of parallel HDF5. There exist some libraries like PNetCDF~\cite{PnetCDF} that bring parallel I/O support to older versions of NetCDF. However, the efficiency of the parallel I/O continues to be heavily dictated by metadata handling and the file system being used.

\subsection{Cinema database generator}
\begin{figure}
  \centering
  \includegraphics[width=\linewidth]{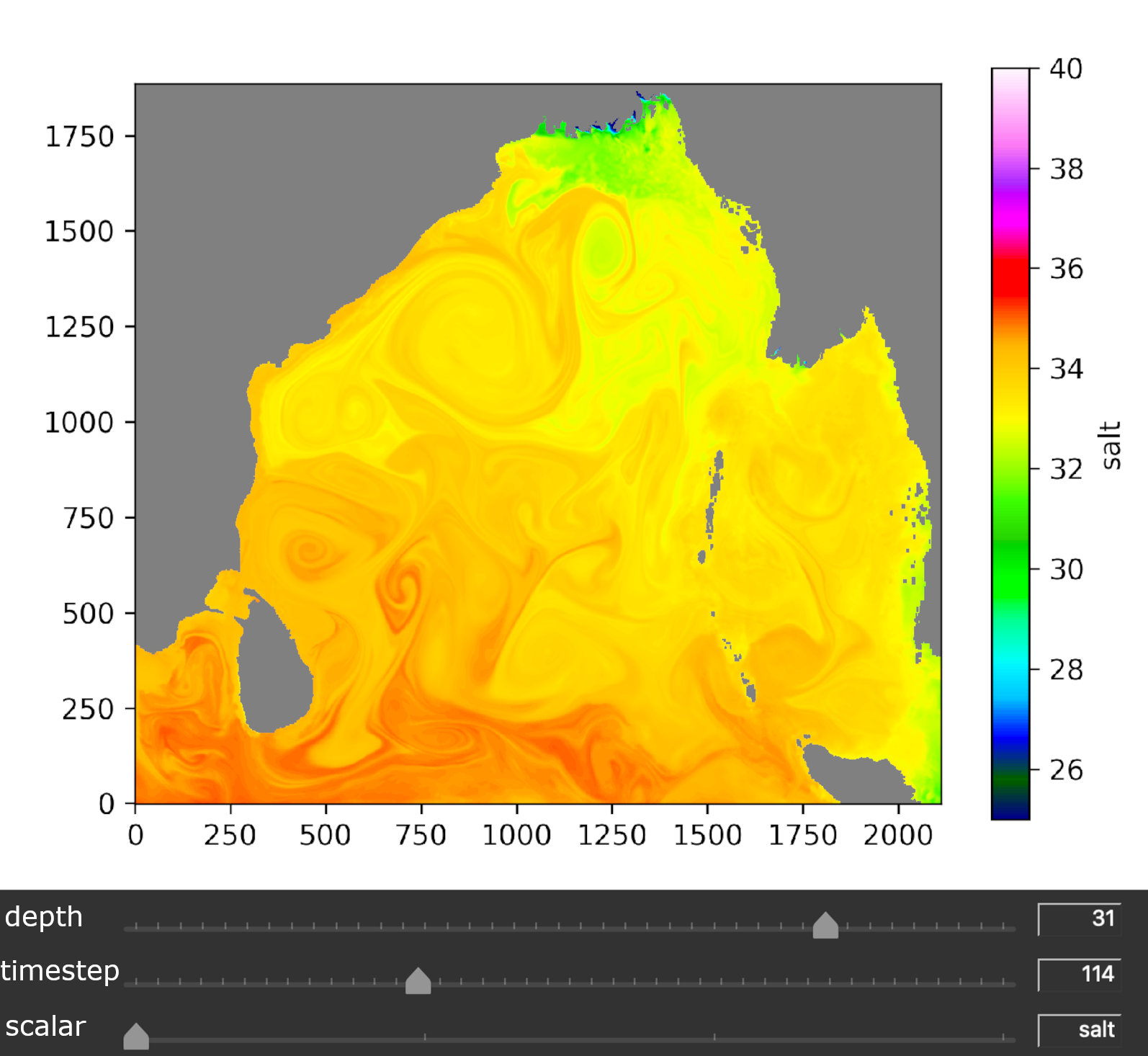}
  \caption{Cinema view with sliding toggles to scroll through the depth slices, time steps, and different scalar fields}
  \label{fig:cinema}
\end{figure}
High-resolution datasets consume a lot of storage space and are impractical to store within the local workstation. The bottleneck of loading the dataset into memory is further amplified due to the additional task of retrieving the dataset from a remote location. The I/O is dependent on the bandwidth of the connection, the distance between the server and the local machine, and the I/O speeds of the storage drives in the server. To circumvent this, we generate an auxiliary dataset that is smaller than the original dataset by several orders of magnitude. One such approach for data reduction and storing data artifacts is the Cinema project, which stores visualizations in an image database for post hoc interactive visualization and exploration of the data. This is especially useful when working with an ocean dataset if the images are generated for every depth slice and every time step. Visualizations in oceanography have traditionally been restricted to 2D depth slices of the ocean. We generate high-resolution float images of the required time steps for every depth slice. This strategy preserves the submesoscale features.

\revision{The cinema database helps the oceanographer swiftly scroll through the data, where the common practice is to study individual depth slices. We note that the cinema generator is flexible enough to allow the user to store the vertical slices instead in the cinema format, as necessary.} An example can be seen in Figure~\ref{fig:cinema}, which shows different sliding toggles to scroll through different depth slices, timesteps, and various scalar fields like salinity, temperature, and velocity. pyParaOcean generates the image database using float images. These images are stored in the standard PNG format, where each pixel contains the corresponding value of the scalar data. This representation enables the user to visualize derived fields within the Cinema viewer directly. The dramatic reduction in I/O times, computation, and storage requirements make it possible for an oceanographer to swiftly scrub through the dataset and identify regions of interest without any friction.

\section{Parallel and distributed computation}\label{sec:methodology}
\begin{figure}
  \centering
  \includegraphics[width=\linewidth]{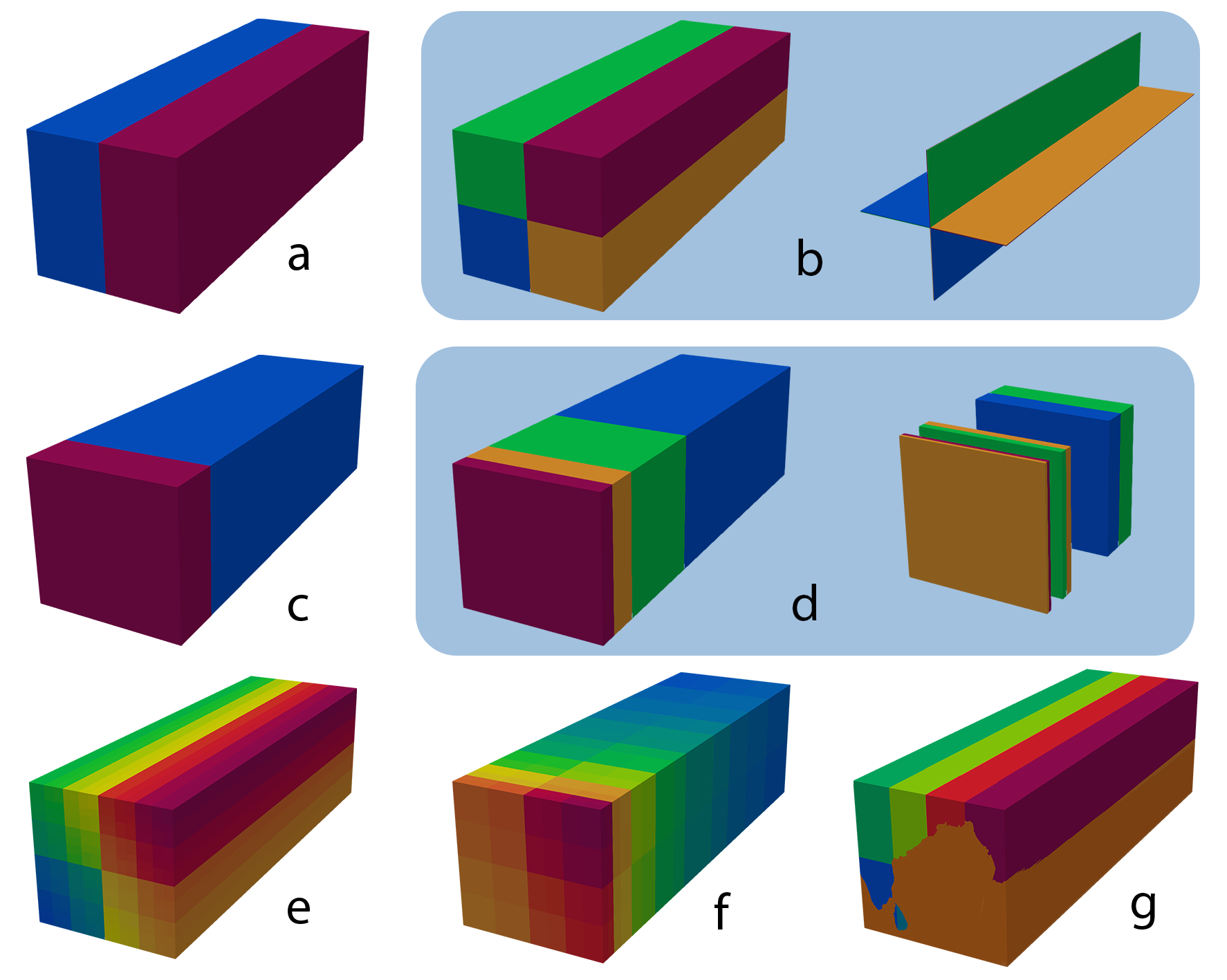}
  \caption{Partitioning the spatial domain of the ROMS dataset for efficient visualization. Each block in the partition, represented using a unique color,  is sent to a unique core that processes the data within the block independently.  (a,c)~partitioning into 2 blocks. (b,d)~partitioning into 4 blocks (left) and the corresponding ghost cells (right). (e,f)~partitioning the domain into 144 blocks. (g)~partitioning the domain into 8 blocks. BoB is shown overlaid in solid orange, indicating that several blocks are restricted to land and hence are not assigned any computational task.}
  \label{fig:partitions}
\end{figure}
In this section, we discuss how pyParaOcean is set up for handling large high-resolution datasets. Distributed parallel computing is enabled in ParaView by launching the pvserver on a remote cluster and connecting to it from the local client. The visualization pipeline is constructed on the client end, and all computation is handled by the remote server. In addition to the size of the dataset, the efficiency is also heavily dictated by how the dataset is distributed across the different cores.

\subsection{Load balancing}
Figure~\ref{fig:partitions} illustrates different approaches to partitioning the spatial domain of the data. Each unique colored block of the partition is processed independently by a different core. The partitions in Figure~\ref{fig:partitions}(a,b,e) are generated automatically by ParaView into 2, 4, and 144 blocks, respectively. In contrast, the partitions in Figure~\ref{fig:partitions}(c,d,f) are the more efficient distribution of data that we propose for this dataset across 2, 4, and 144 blocks, respectively.

The efficiency of the pyParaOcean filters depends on achieving balanced data distribution across processing cores. Blocks containing only the landmass data points correspond to zero computational workload and create bottlenecks due to load imbalance. The ocean region solid orange (\inlinegraphics{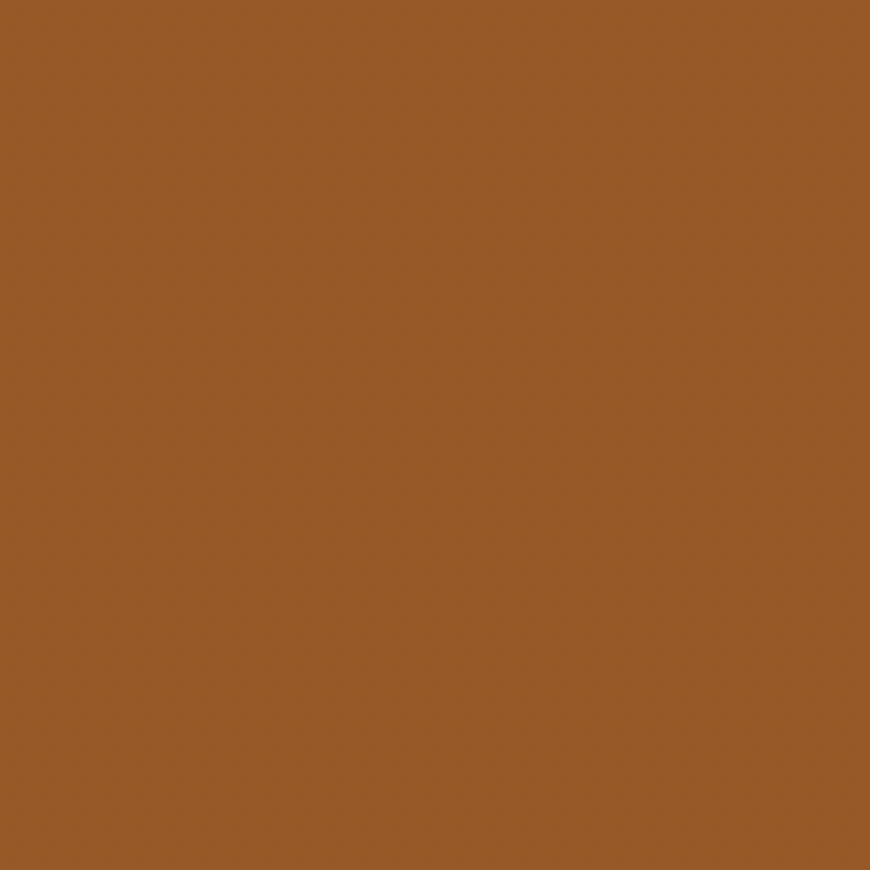}) is overlaid on the grid in Figure~\ref{fig:partitions}(g) to help identify the blocks that contain ocean data points. The two green blocks (\inlinegraphics{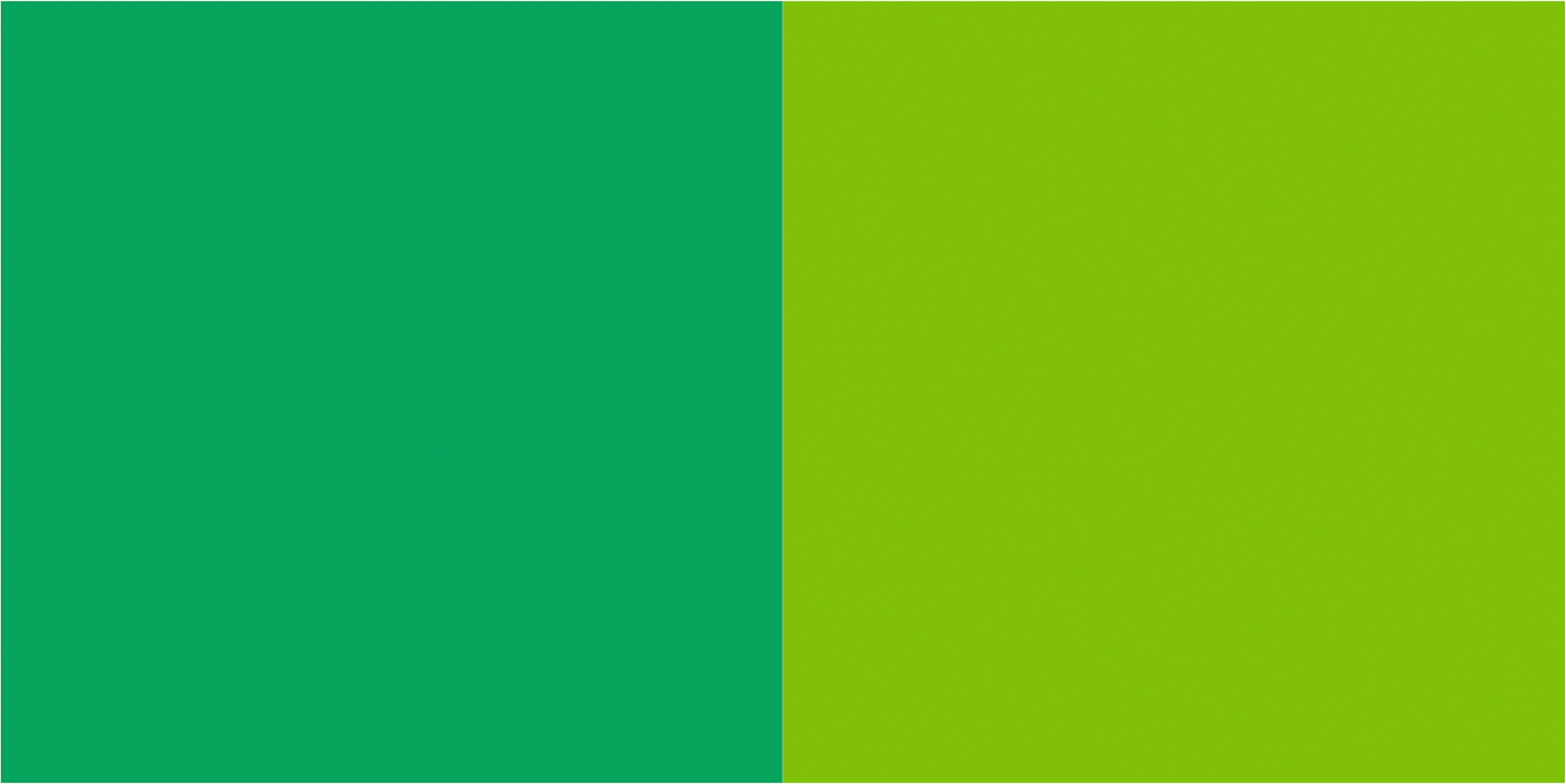}) on the top left corner in Figure~\ref{fig:partitions}(g) do not overlap with the ocean data points. These blocks consist only of landmass data points, which correspond to NaN values, and do not contribute to the computation. Similarly, in Figure~\ref{fig:partitions}(b) the top-left green block (\inlinegraphics{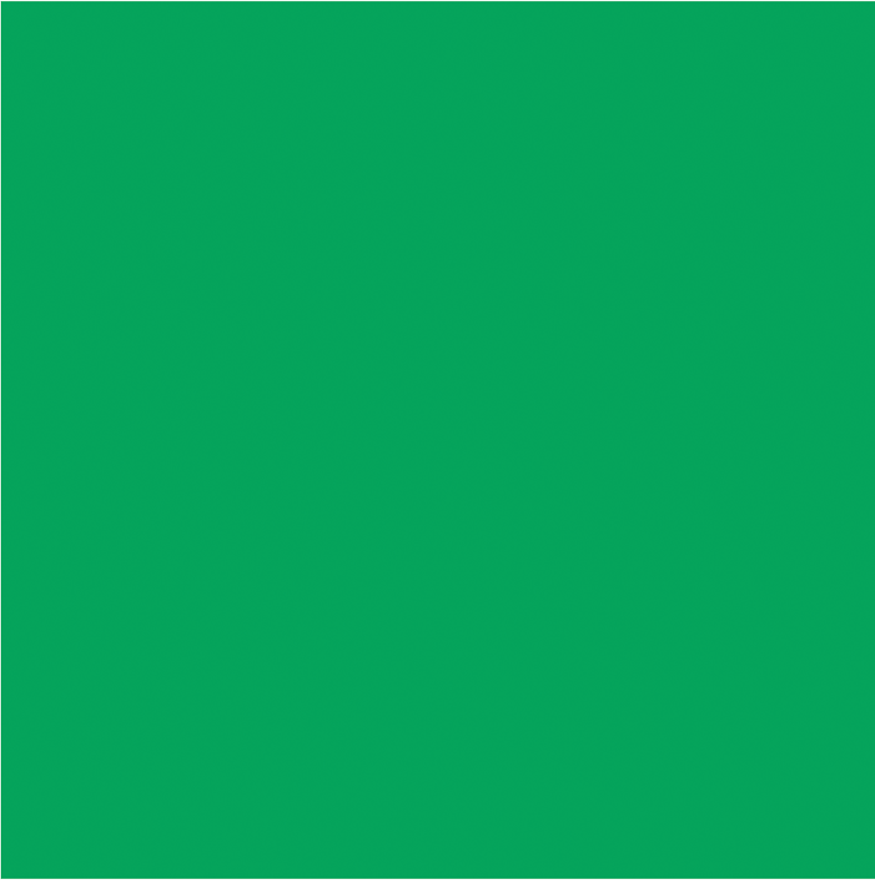}) contains only landmass data points, while the other three cores perform more computations since they receive actionable ocean data points. Hence, Figure~\ref{fig:partitions}(d) is a better partition which assigns equal amount of data to all four cores. Further, the ocean data is distributed equally between the four cores, thereby eliminating the load imbalance. Similarly, the partition in Figure~ \ref{fig:partitions}(c) is better than the one shown in Figure~\ref{fig:partitions}(a) because the blue block (\inlinegraphics{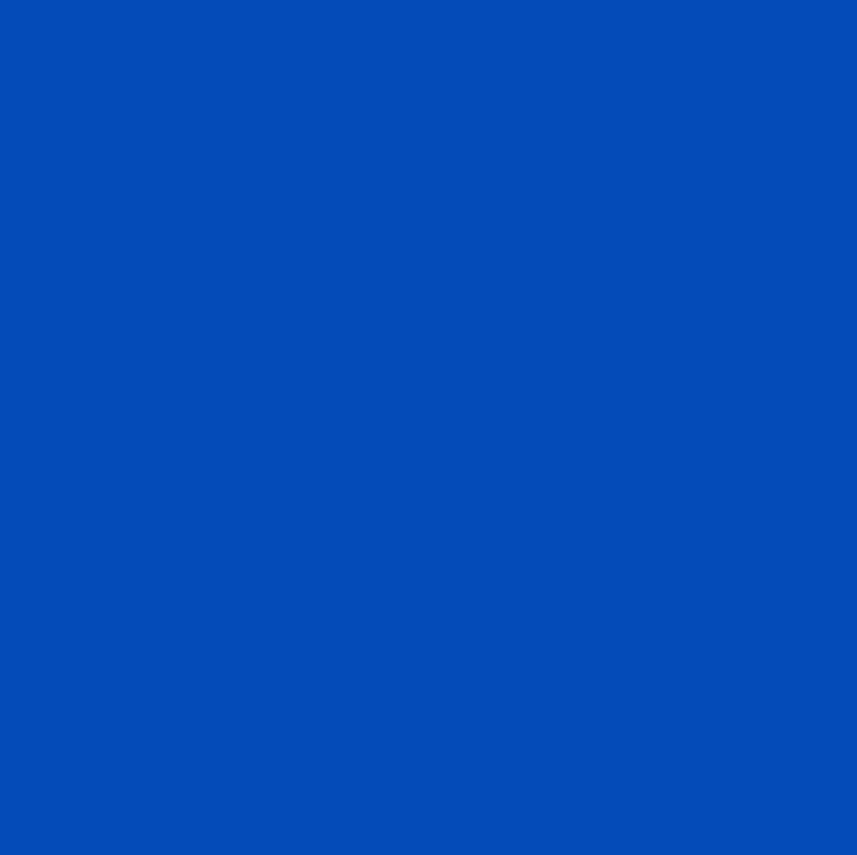}) contains fewer ocean data points as compared to the red block (\inlinegraphics{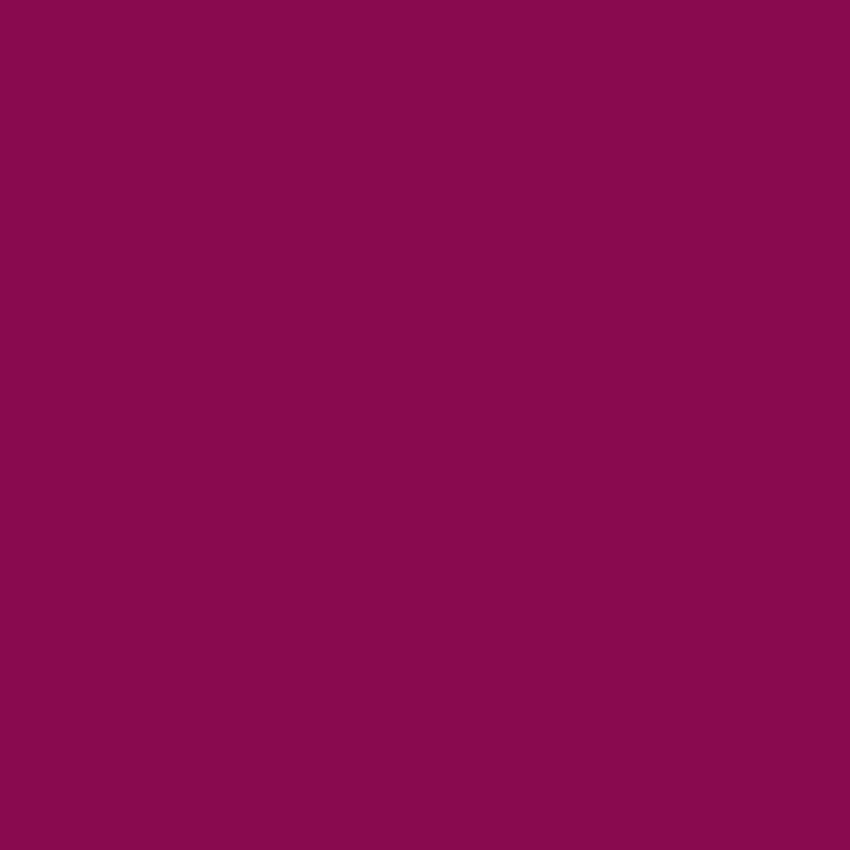}).

Filters such as the fieldline generator do not require interaction between data items in the vertical (depth) direction. In this case, partitioning the data along the depth slices offers a significant advantage.  Since the computation is performed independently within a slice, the filter does not require to access information from neighboring slices. In contrast, partitioning along latitude or longitude results in a block that requires communication with neighboring blocks via ghost cells for streamline computation. Data is requested through ghost cells only on demand. So, the ghost cells in Figure~\ref{fig:partitions}(d) are not used to compute the output of the fieldline computation filter, whereas the ghost cells in Figure~\ref{fig:partitions}(b) are used and the filter requests data from them. So, even though the number of ghost cells in the former case is considerably larger, the partition remains computationally efficient because the ghost cells are not utilized.  The ocean data has lower resolution along the depth direction. So, the partition has to eventually be along latitude and longitude. So, with increasing number of blocks, the computation times for the two partitioning strategies become similar. The similarity between the partitioning schemes is visible from Figures \ref{fig:partitions}(e,f).

\subsection{Parallel front tracking}\label{subsec:MFT}
A  recently developed front-based method for tracking of salinity movement has helped document HSC propagation in the BoB~\cite{singh2022frontskeleton}. This algorithm for front computation and tracking is now implemented in parallel across time steps and across individual depth slices, resulting in an efficient pyParaOcean module. 

Front-based tracking proceeds in two major steps. First, extract an isovolume ($\geq$ 35 psu) and compute the boundary curve as the intersection of the isovolume with each depth level. The boundary curve may consist of multiple components. Segment each connected component of the boundary curve into a north-facing segment and a south-facing segment. Next, connect such north-facing segments across all depth levels if they lie within a small neighborhood specified by a distance parameter $n$,  resulting in surface fronts. In the second step, the surface front correspondence between time steps is identified using the parameter $n$. Two fronts correspond from consecutive time steps correspond to each other if they lie within a small spatial neighborhood specified by the distance parameter $n$ or spatially overlap with each other.  A track graph, whose vertices correspond to the surface fronts and whose directed arcs correspond to pairwise connections between surface fronts, is constructed to summarize all possible movements of the front. 

In the first step of the method, all computations within a time step are independent of other time steps. Hence, they are executed in parallel. In the second step, computing arcs between two consecutive time steps requires that all surface fronts within the two time step are already computed. This necessitates a synchronization at the end of the first step. The neighborhood search and correspondence computation is sped up by representing the isovolume as a binary grid and utilizing simple matrix operations. The boundary of the isovolume is computed using a simple $3\times 3$ averaging filter, followed by a selection of voxels whose value lie strictly between 0 and 1. This boundary is further processed to extract the north-facing boundary, which is eventually stored again as a binary 3D grid (1s representing the boundary). 
Neighbor search for a given distance parameter $n$ is made efficient by transforming it into a simple overlap problem. Each 1 is expanded into a $n \times n \times 2$  neighborhood of 1s centered at the voxel and extending to the next time step. A subsequent connected component labeling step using a 26-neighbor connectivity labels individual connected components of the expanded grid of 1s. Multiplying the values in this grid with the corresponding values of the boundary grid produces the grid containing the surface fronts, where each voxel has a unique label that indicates the component of the surface front. 

Each component of the surface front corresponds to a vertex of the track graph. We use an $n \times n$ mask that consists of 1s for voxels in the circular neighborhood of radius $n$ and 0s otherwise. Arcs of the track graph are computed using this circular binary mask of radius $n$ in time step $t+1$, centered at all voxels whose label is non-zero in time step $t$. For each such voxel with a non-zero label in time step $t$, we multiply the mask with the grid value at the voxel in time-step $t+1$. All values outside the neighborhood are set to 0, resulting in a collection of unique non-zero labels. These labels are unique identifiers of surface front components, where each label represents the destination of the arc and the origin of the arc is the surface front containing the voxel from time step $t$. 
This implementation results in significant runtime improvements because several of the computations are transformed into matrix operations. 

The method described above is executed in parallel for each time steps. The computation is parallelized over all available cores using MPI or the Python multiprocessing library. A process is responsible for a single time step $t$. The process computes surface fronts within time step $t$, and waits until surface fronts for time step $t+1$ are computed before computing arcs between the two time steps. The total runtime is dominated by the longest running process, provided that a sufficient number of cores are available to independently handle each time step. Moreover, all the grids and maps are stored as NumPy arrays~\cite{harris2020array} and all grid operations are performed using inbuilt NumPy methods. NumPy is faster at grid operations because it supports array processing without requiring to individually address each element. Further, NumPy uses machine-native data types as opposed to Python's object types and provides fast implementations in languages such as C and Fortran for each grid and matrix operation. Many of the NumPy operations also leverage multithreading. The culmination of all these factors results in fast computations.

\section{Scaling studies}\label{sec:experiments}
In this section, we present an experimental study of the scaling behavior of the pyParaOcean modules for computing and visualizing ocean structures. 

\subsection{Experimental setup}
All scaling experiments are performed on a cluster consisting of eight nodes. Each node contains an AMD EPYC Processor with 32 cores and 256GB RAM. For executing the front-based salinity tracking, we use MPI to schedule processes on cores in all cluster nodes. All scripts are written in Python3.

\subsection{pyParaOcean filters}
\begin{figure*}
  \centering
      \subcaptionbox{\label{fig:seedplacementscaling}}{\includegraphics[width=0.45\linewidth]{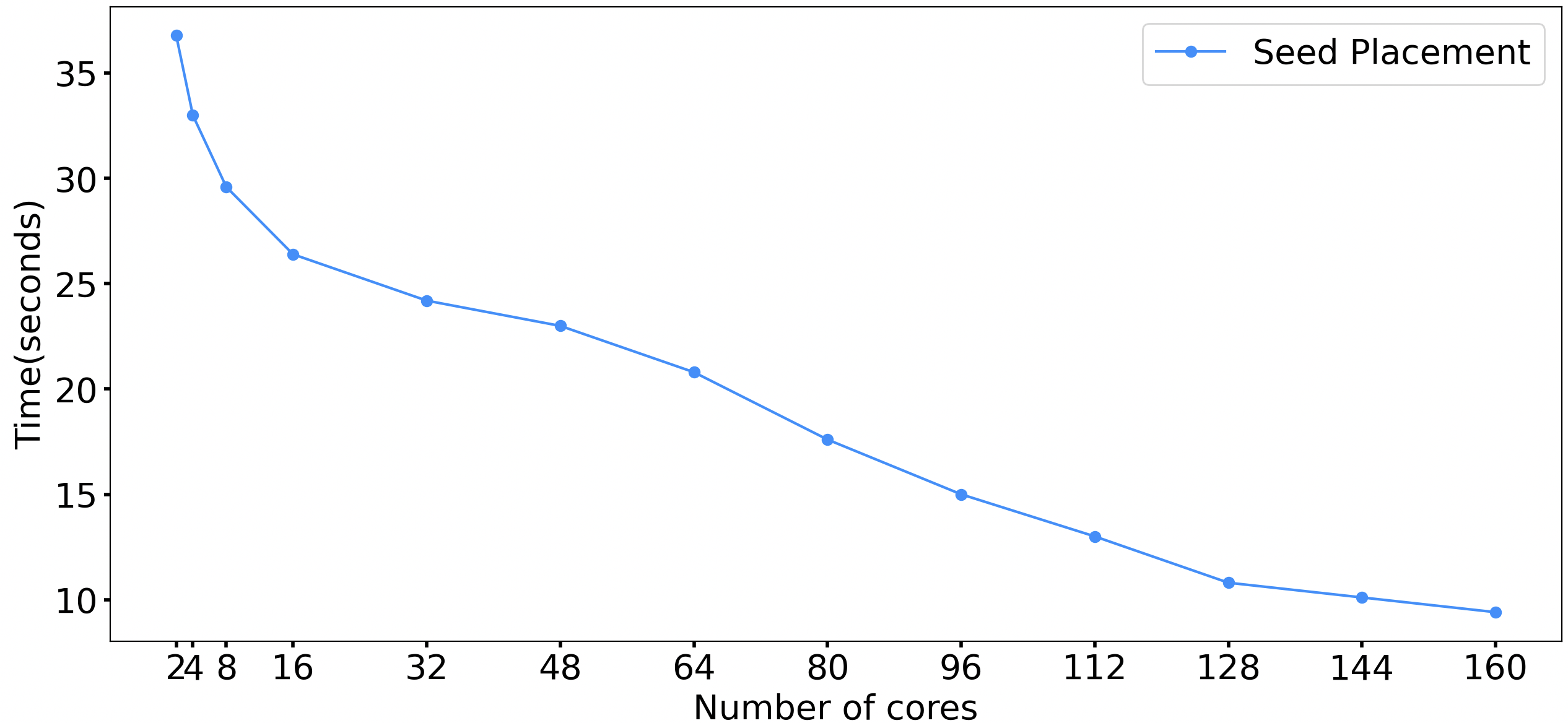}}      \subcaptionbox{\label{fig:streamlinesscaling_differentdatasets}}{\includegraphics[width=0.45\linewidth]{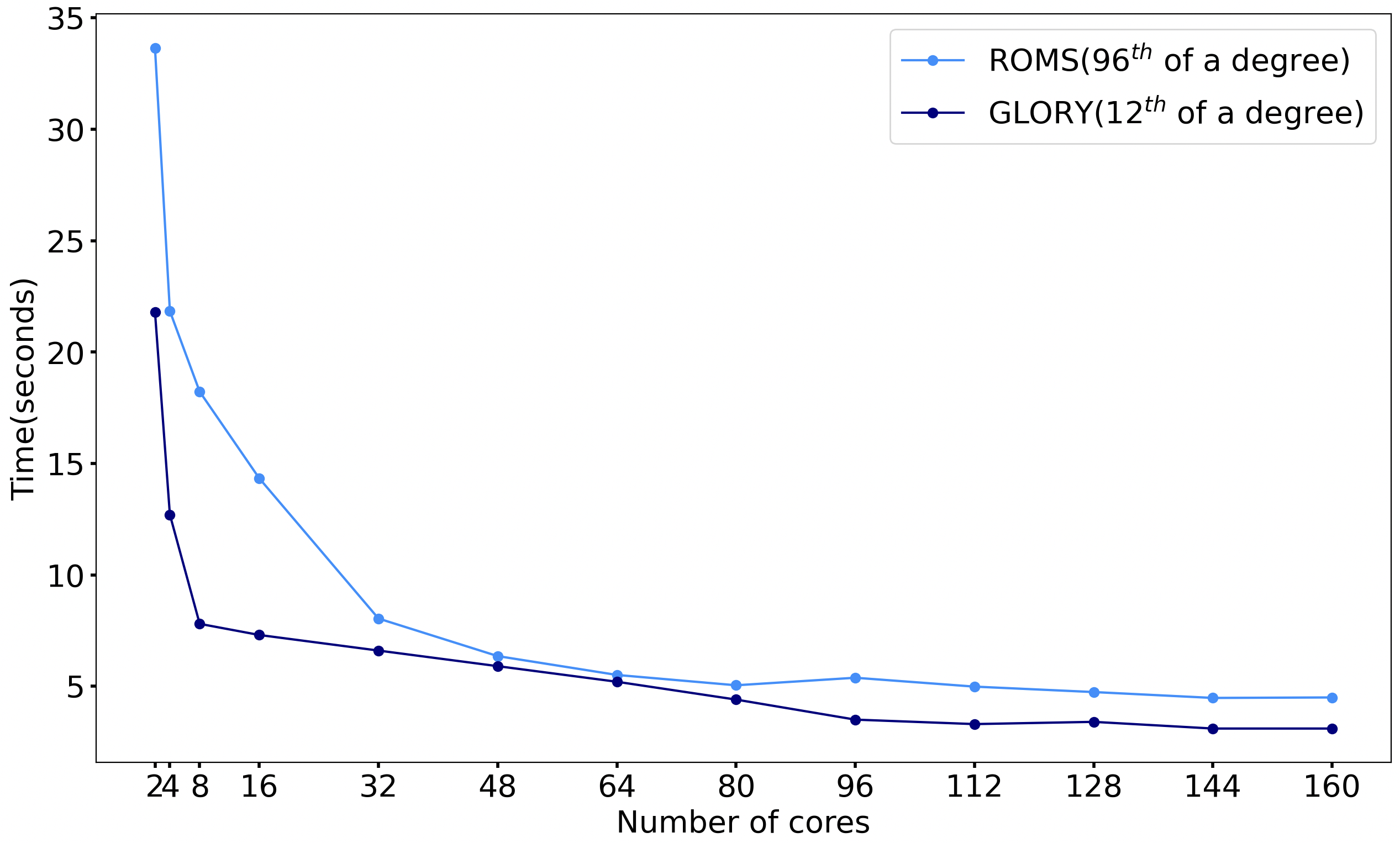}}      \subcaptionbox{\label{fig:pathlinescaling}}{\includegraphics[width=0.45\linewidth]{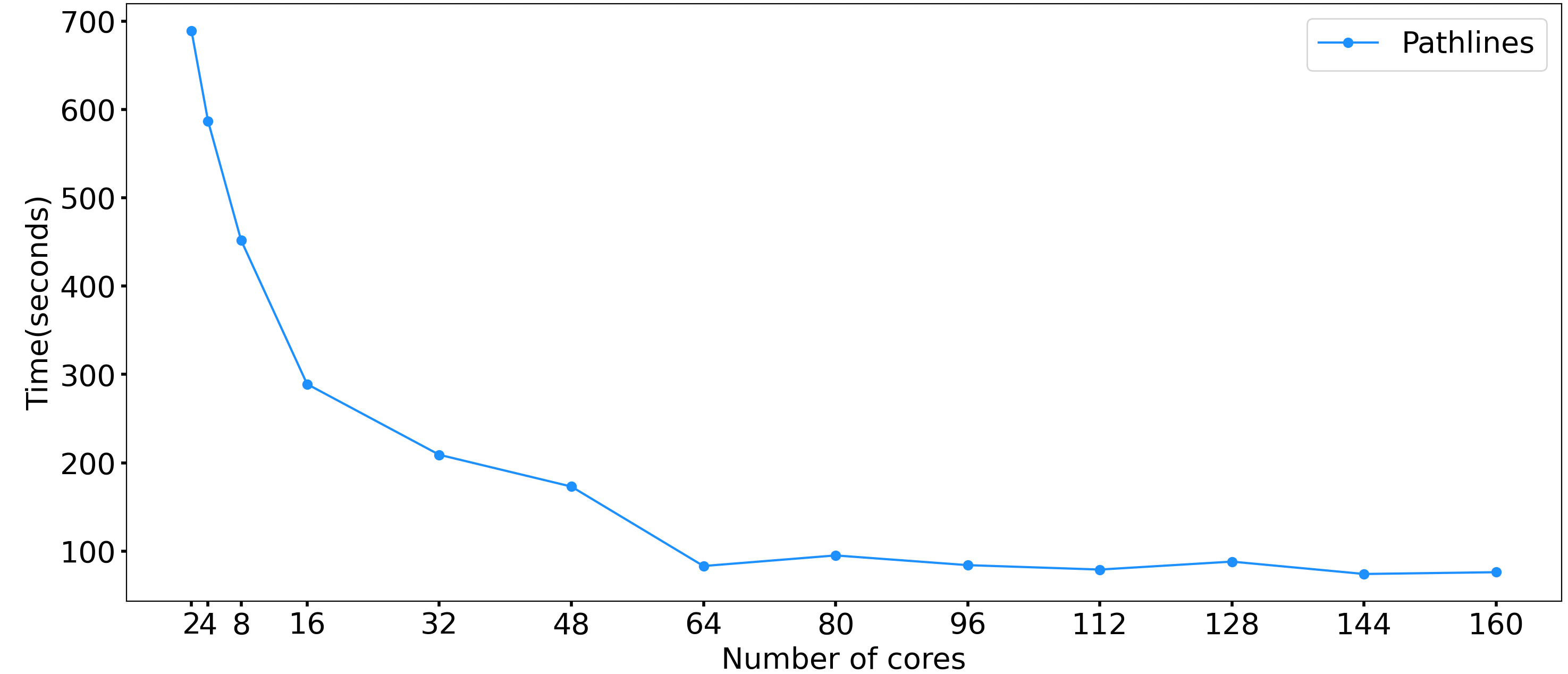}}      \subcaptionbox{\label{fig:isovolumescaling}}{\includegraphics[width=0.45\linewidth]{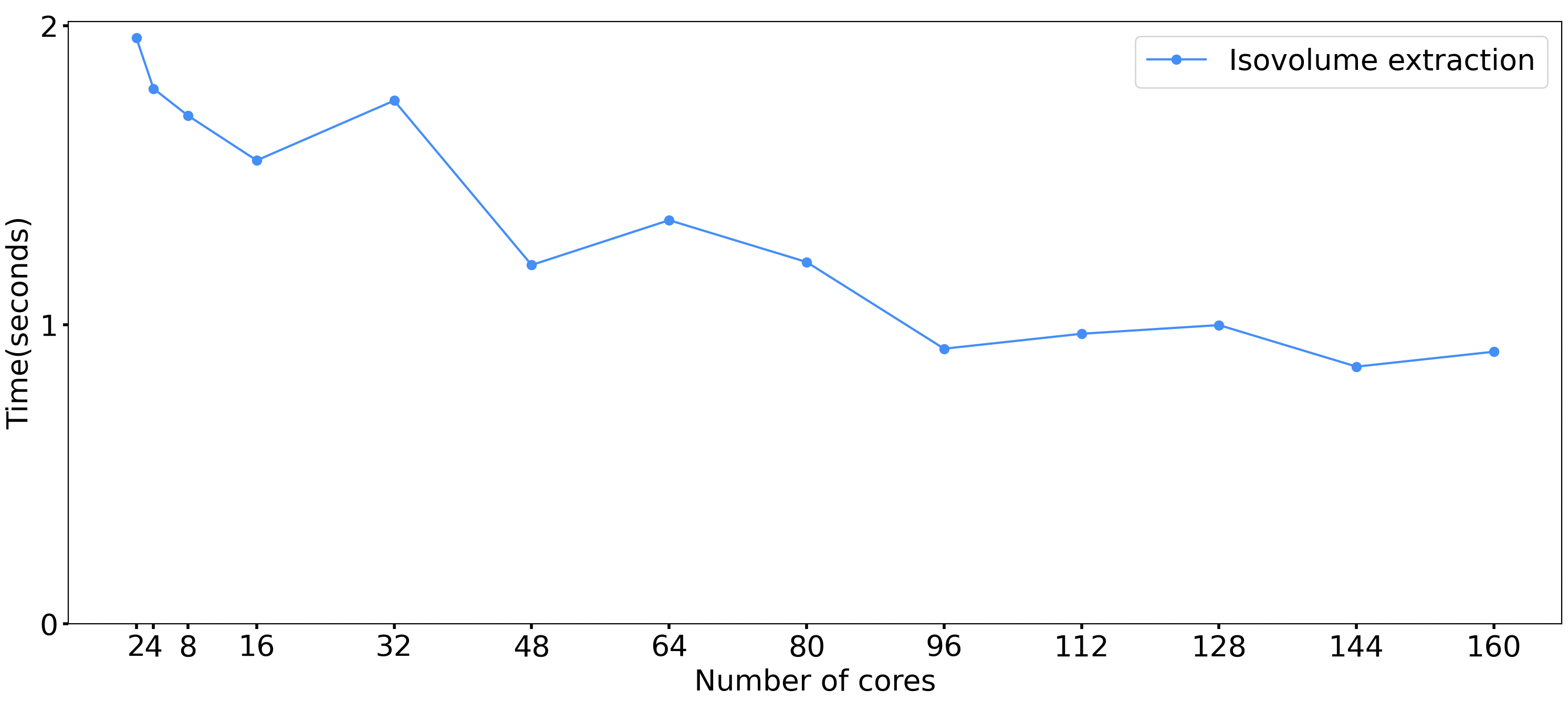}}
  \caption{Scaling behavior of pyParaOcean filters. (a)~The seed placement filter applied on the ROMS dataset. The seeds are weighted according to vorticity. The scaling is linear, with a steeper drop in runtime up to 16 cores followed by a smaller slope. (b)~The streamlines filter computed on 500 seeds. Runtimes drop initially for both ROMS and GLORYS datasets. After the initial steep drop, the plot saturates at 80 cores. (c)~The pathline filter applied on 500 seeds, particles tracked over 10 time steps. After an initial sharp drop, runtimes saturate past 64 cores. (d)~Isovolume extraction filter applied on the ROMS dataset. Runtimes steadily decrease but saturate at 96 cores. The fluctuations in the runtime may be attributed to the I/O times of the logger routine.}
\end{figure*}
The filters in pyParaOcean demonstrate excellent scalability with increasing number of processing cores. This allows for efficient handling of large and complex ocean datasets. Figure~\ref{fig:seedplacementscaling} shows the scaling behavior of the seed placement tool with increasing number of processing cores. We load the dataset using the default partitioning scheme of ParaView and execute the filter to place 1000 seeds weighted according to the vorticity field. This ensures that a larger number of seeds are placed in regions with high vorticity.  The calculation of the vorticity field and subsequent placement of seeds is parallelized. We observe that the runtime reduces from 36s across two cores to approximately 9s across 160 cores. A similar trend is observed when we execute the streamline (Figure~\ref{fig:streamlinesscaling_differentdatasets}) and the pathline (Figure~\ref{fig:pathlinescaling}) filters. Both filters are executed with 500 seeds. Figure~\ref{fig:streamlinesscaling_differentdatasets} shows the runtimes, both on the ROMS and the GLORYS datasets. We observe that the filter scales well, and saturates at approximately 80 cores,  after which there is no significant speedup. A similar saturation is also observed in Figure~\ref{fig:pathlinescaling} where the speedup for the particle tracer saturates after approximately 64 cores. The runtimes reported in Figure~\ref{fig:pathlinescaling} are based on an experimental run that preloads 10 time steps into the memory. The particle tracer filter runs on multiple time steps and each time step in the ROMS dataset is stored as a different NetCDF file in a server. Therefore, to report the correct computation times that are not inflated due to I/O and the network bandwidth constraints, we preload these files into memory before beginning the computation. We discuss I/O in greater detail within Section~\ref{subsec:io}.

The isovolume provides a useful overview but is computationally heavy due to the size of the data that is processed and produced as output.  Figure~\ref{fig:isovolumescaling} shows a general downward trend in the runtime with increasing number of cores. The total runtime of the filter is less than 2s. The fluctuations in the runtimes are likely due to wait times when multiple processes attempt to write into the same log file.

\subsection{Parallel front tracking}
\begin{figure*}
  \centering
      \subcaptionbox{\label{fig:roms1}}{\includegraphics[width=0.45\linewidth]{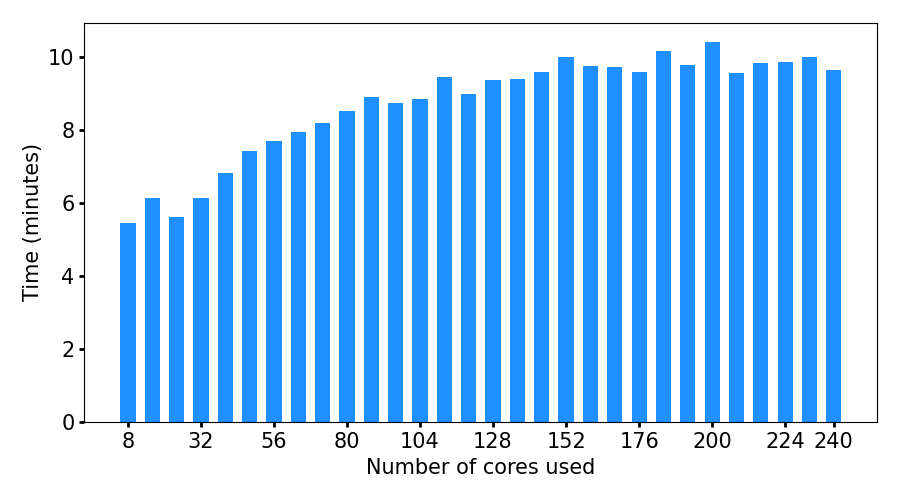}}  
      \subcaptionbox{\label{fig:cn1}}{\includegraphics[width=0.45\linewidth]{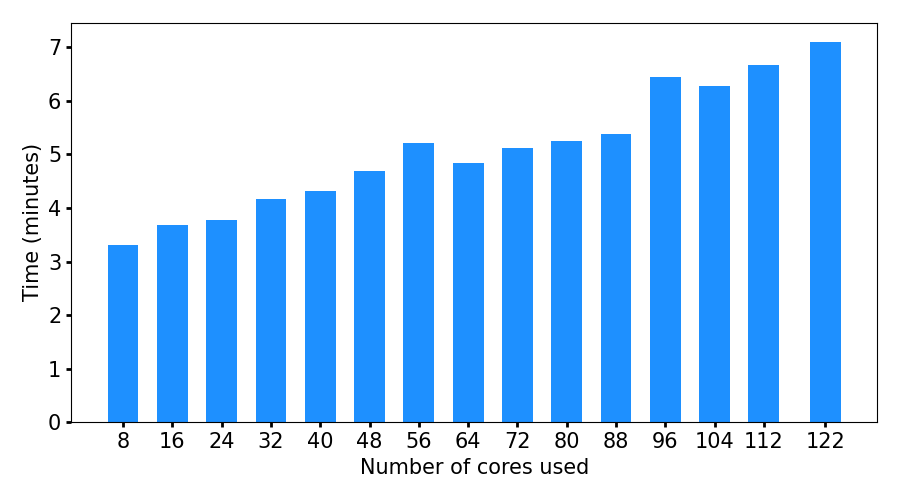}}  
       \caption{Weak scaling study. (a)~ROMS data with a resolution of 96$^{th}$ of a degree. (b)~GLORYS data with a resolution of 120$^{th}$ of a degree. In both datasets, we observe an increase in runtime as the number of cores and time steps increase. The curve appears to flatten towards the end for the ROMS dataset.}
\end{figure*}
\begin{figure*}
  \centering
      \subcaptionbox{\label{fig:roms3}}{\includegraphics[width=0.45\linewidth]{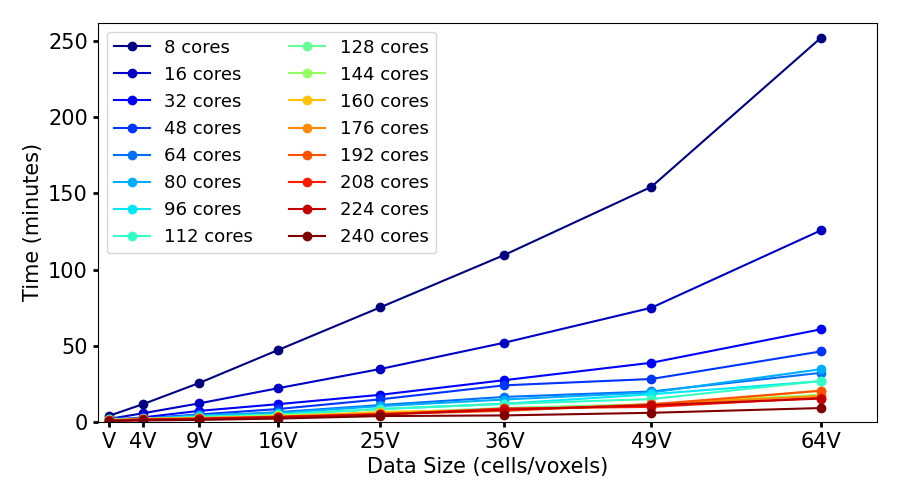}}  
      \subcaptionbox{\label{fig:cn3}}{\includegraphics[width=0.45\linewidth]{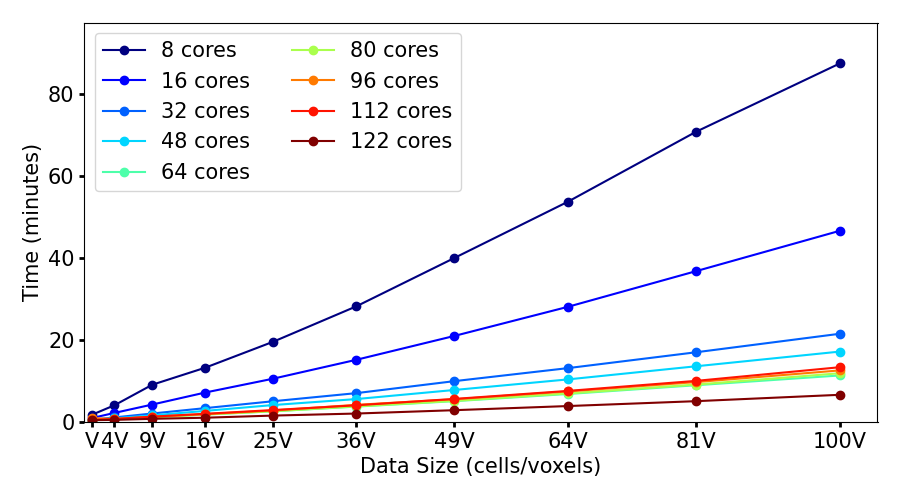}}  
       \caption{Runtime complexity. (a)~Processing 240 time steps of increasing resolutions of the ROMS data using multiple cores. Let V denote the data size at a resolution of 12$^{th}$ of a degree, which corresponds to $265\times 229 \times 200$ voxels at each time step or $265\times 229 \times 200 \times 240$ voxels in total. (b)~Processing 122 time steps of increasing resolution of the GLORYS data using multiple cores. Again, V denotes the data size at a resolution of  12$^{th}$ of a degree, which in this case equals $253\times 301 \times 200$ voxels at each time step or $253\times 301 \times 200 \times 122$ voxels in total. In both datasets, runtime increases linearly with data size. We observe a spike at 96$^{th}$ of a degree (64V) for the ROMS dataset.}
\end{figure*}
\begin{figure*}
  \centering
      \subcaptionbox{\label{fig:roms2}}{\includegraphics[width=0.45\linewidth]{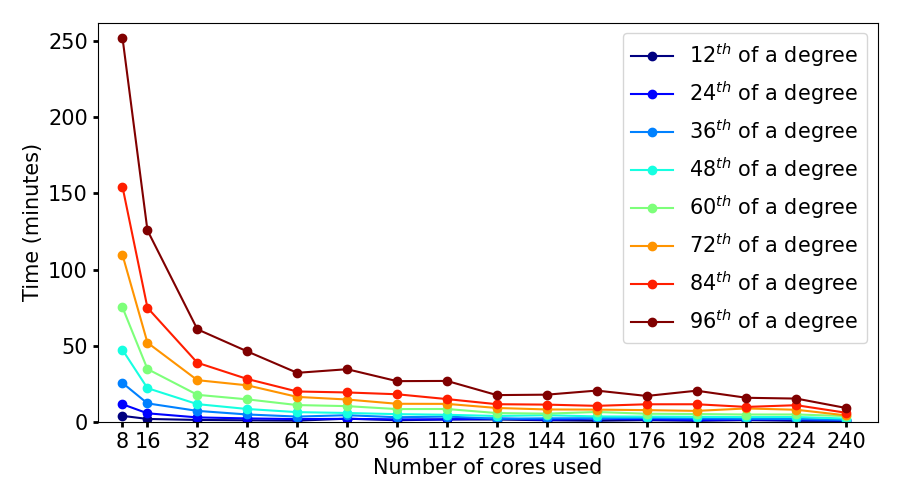}}  
      \subcaptionbox{\label{fig:cn2}}{\includegraphics[width=0.45\linewidth]{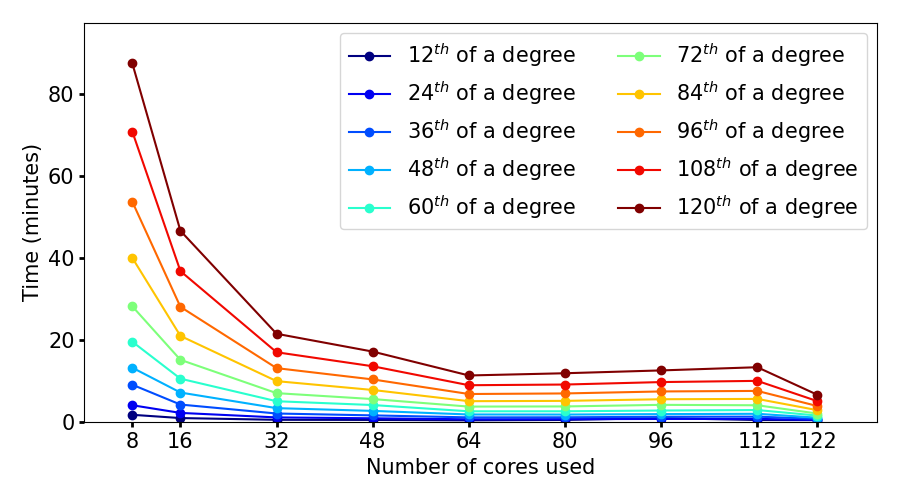}}  
       \caption{Strong scaling study. (a)~Processing 240 time-steps of a fixed resolution ROMS data using an increasing the number of cores. (b)~Processing 120 time steps of a fixed resolution GLORYS data. In both cases, runtime drop is steep initially followed by a eventual flattening of the curve.}
\end{figure*}
Section~\ref{subsec:MFT} described a parallel algorithm for front-based tracking of the HSC.  In the ideal scenario, if the number of available cores is equal to the number of time steps and all cores are utilized then the run time should remain constant with increasing number of time steps. Providing a larger number of cores than the number of time steps is wasteful for the current implementation of the algorithm. In our first experiment, we evaluate this weak scaling behavior. We observe an increase in runtime even as the number of cores is increased in tandem with the number of time steps, as seen in Figures~\ref{fig:roms1} and~\ref{fig:cn1}. This increase is not large, is within expectations, and the curve seems to flatten as the number of time steps is increased further. This weak scaling behavior is indicative of the applicability of the implementation to larger time periods. 
 
The front-based tracking algorithm is executed on the ROMS and GLORYS data at multiple spatial resolutions to study the runtime complexity in practice. Executing 240 time steps at a high resolution requires more RAM than available in our experimental setup. On average, it takes 9.6 and 7.1~minutes to process the largest volumes of ROMS ($2113\times 1825 \times 200 \times 240$ voxels) and GLORYS ($2521\times 3001 \times 200 \times 122$ cells/voxels), respectively. 

The aim of the next experiment is to study the runtime complexity. A linear increase in resolution across latitude and longitude results in a quadratic increase in the total size of the data. In the ideal scenario, we expect the runtime to increase linearly with the data size, given that the computational power (number of cores) remains constant. Figures~\ref{fig:roms3} and~\ref{fig:cn3} show that our implementation exhibits close to linear scaling suggesting its applicability for larger datasets with predictable run times. Figure~\ref{fig:roms3} shows a spike in runtime at a spatial resolution of $96^{th}$ of a degree. We believe this sharp increase in runtime is due to the size of data approaching the upper limits of the available RAM. We restrict the experiments on ROMS to this spatial resolution due to the memory limitation.

For a fixed data size, increasing the number of computational cores results in an improvement in run time as shown in Figures~\ref{fig:roms2} and~\ref{fig:cn2}. This strong scaling behavior is studied in our final experiment. The improvement in runtime flattens eventually as discussed earlier.

\subsection{Effect of data distribution}
\begin{figure*}
  \centering
      \subcaptionbox{\label{fig:ghostcellsgeneration}}{\includegraphics[width=0.51\linewidth]{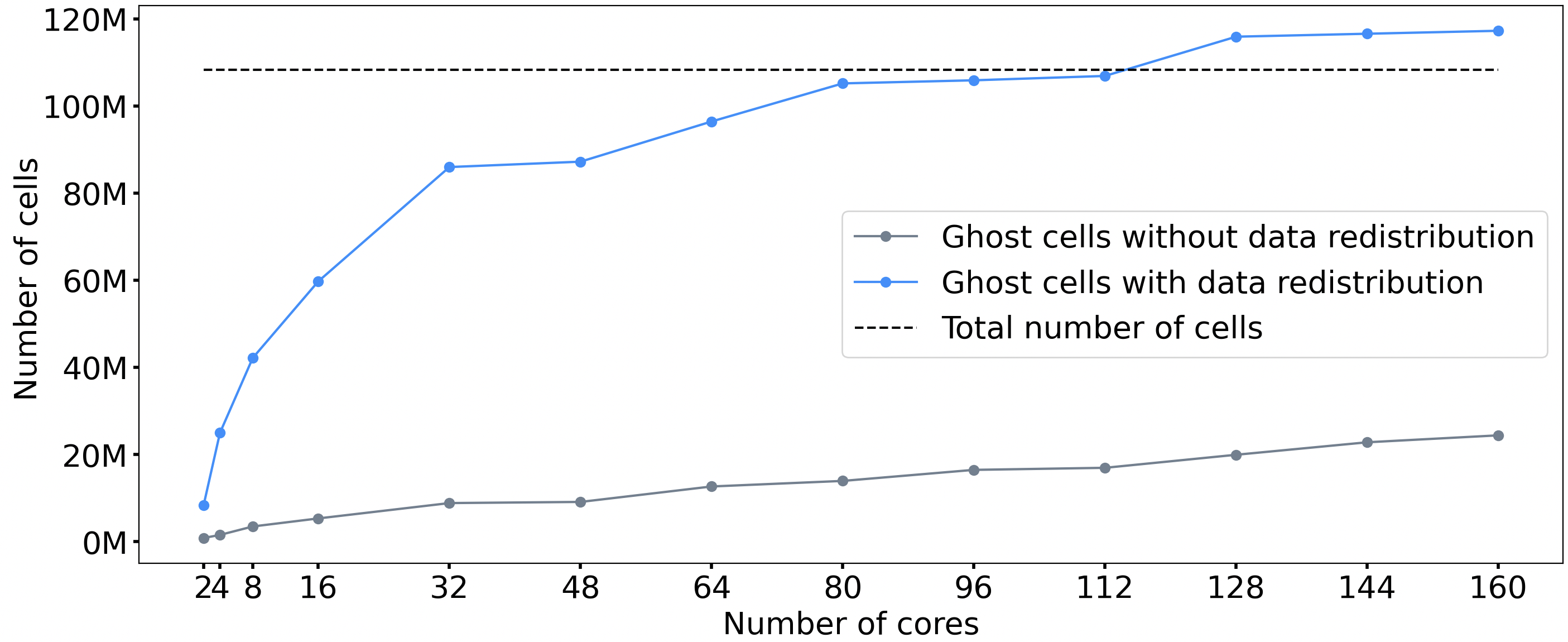}}  
      \subcaptionbox{\label{fig:scalingstreamlinesd3}}{\includegraphics[width=0.42\linewidth]{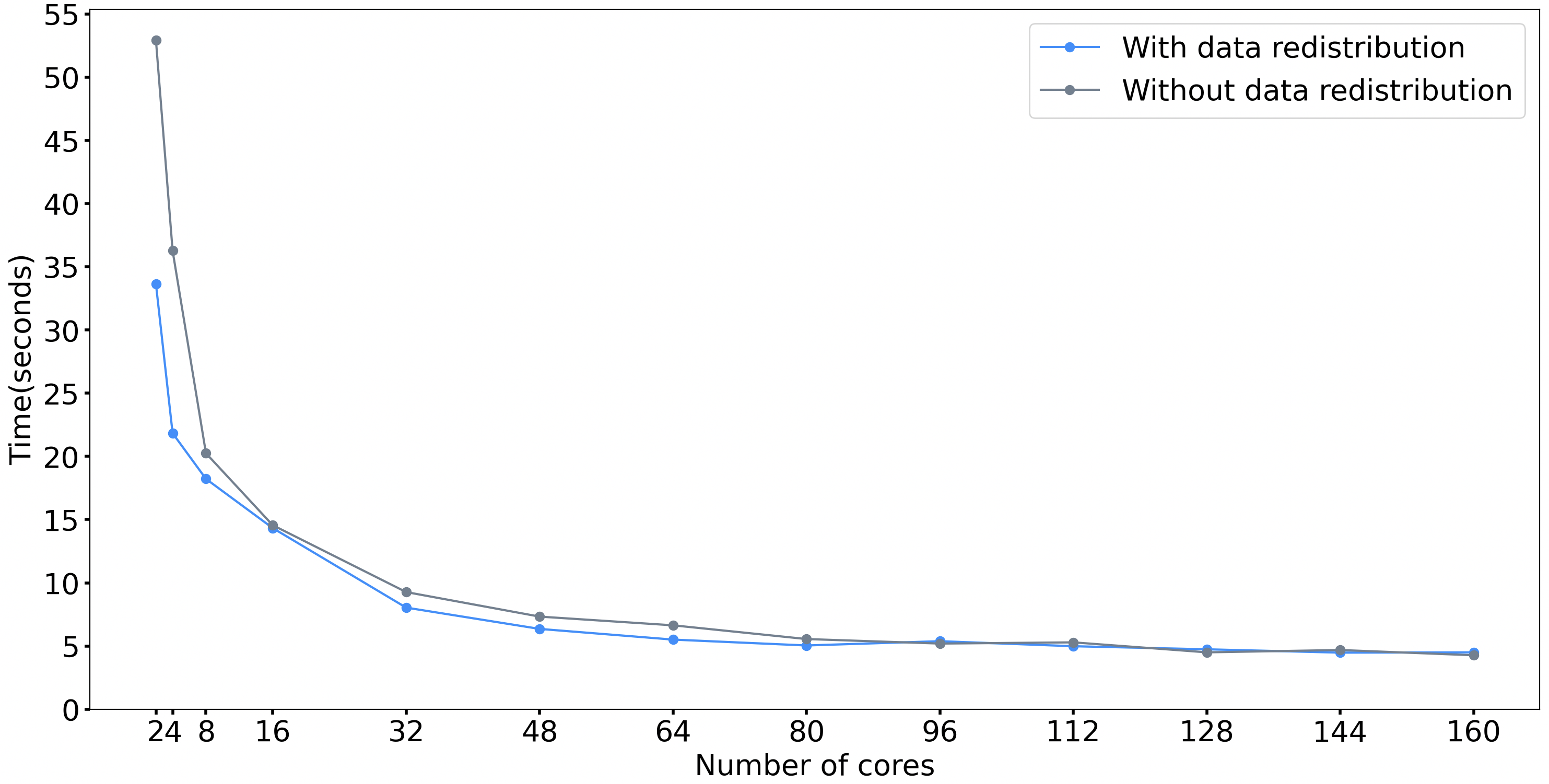}}  
       \caption{Effect of data distribution on scaling behavior. (a)~A larger number of ghost cells are generated after data redistribution. (b)~Streamlines are integrated with 500 seeds in the ROMS dataset. The filter is significantly \revision{faster for a smaller number of cores} after redistribution, even when the number of ghost cells is large. But, runtimes are similar for larger number of blocks when the distributions become similar as shown in Figure~\ref{fig:ghostcells}.}
\end{figure*}
The scalability of the filters in ParaView is highly dependent on how the data is distributed across the processing cores. Figure~\ref{fig:scalingstreamlinesd3} shows the runtimes for the streamline filter with and without redistribution. The redistribution offers a significant advantage for smaller number of cores (2,4,8). The runtime improvement may be attributed to two reasons. Firstly, as we can see from Figures~\ref{fig:partitions}(a,b,g), the default partition skews the data distribution balance. Some nodes are assigned blocks corresponding to land, which does not correspond to any computation. Further, streamline computation requires information from neighboring blocks since the data is partitioned along latitude and longitude. This adds an overhead of requesting and computing additional information from the neighboring blocks in the form of ghost cells. Both drawbacks are eliminated by slicing the data along the depth dimension, as observed from Figures~\ref{fig:partitions}(c,d). First, the skew in the data distribution is eliminated by cutting along depth because each process is assigned an equal amount of land and ocean data. Second, streamline computation does not utilize vertical velocity. \revision{Generally, vertical velocities in the ocean are several order of magnitudes smaller than the horizontal velocities. The effect of vertical velocities is negligible for visualization of streamlines, which essentially represent the horizontal flow field. Hence, the processes do not require the ghost cells. The exceptions are eddies and small scale turbulence.}

Figure~\ref{fig:ghostcellsgeneration} plots the number of ghost cells generated without redistributing the data (grey) and with data redistribution (blue). The black dashed line indicates the total number of voxels in the dataset. Even though the number of ghost cells is considerably higher with redistribution as compared to the default partition from ParaView, the runtimes with redistribution are lower. However, as the number of processing cores increases, the two partitioning schemes become similar, see Figures~\ref{fig:partitions}(e,f)). Hence, the runtimes for both partitioning schemes also converge to similar numbers.

\subsection{I/O and the Cinema database}\label{subsec:io}
\begin{figure}
  \centering
  \includegraphics[width=\linewidth]{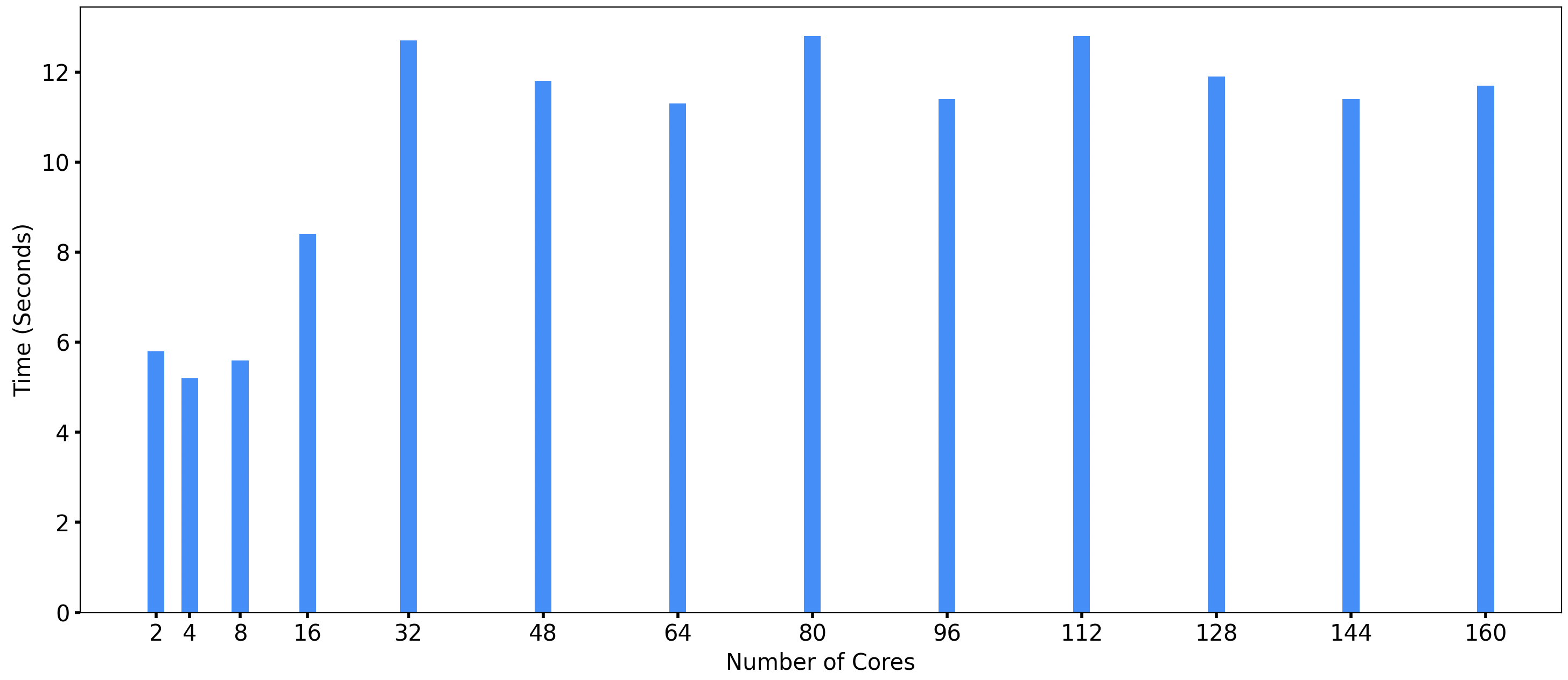}
  \caption{I/O time increases with number of cores. Time taken to load a single time step in the NetCDF format from the ROMS dataset. The time to load a file into memory increases with number of cores and flattens after 32 cores.}
  \label{fig:io}
\end{figure}

Figure~\ref{fig:io} shows the average time taken to load a single time step into memory. NetCDF is a self-describing file format. Both metadata and the actual data are stored within the same file. A file reader is expected to read the metadata to understand the data format. A consequence of this feature is that when multiple processes are launched to execute a filter, all the processes attempt to read the common metadata. While the data is distributed among the processes and can be read with some degree of parallelism, reading the common metadata causes a serialization. Hence, as the number of cores increase, we observe an increase in the time taken to load the dataset into memory. Another factor that impacts the time to load a file is the network bandwidth. For large data sizes, the network bandwidth could become the bottleneck, overshadowing any improvements in the I/O speeds. So, it becomes impractical to load a time step to render a quick overview for the user. 

In order to support the ability to provide a quick overview of the data, we propose the generation of a Cinema database~\cite{ahrens2014image} for local storage. We used our Cinema generator to generate float images for 4 scalars namely salinity, temperature, velocities in $x$ and $y$ directions for a total of 100 time steps. This resulted in a reduction in data size from approximately 750GB to 2.6GB. The Cinema database is 0.35\% of the original but captures all attributes that are typically studied by an oceanographer to get an overview of the data. The generator can also be tuned to generate and store additional fields.

\section{Case study}\label{sec:casestudy}
\begin{figure}
    \centering
    \includegraphics[width=0.7\linewidth]{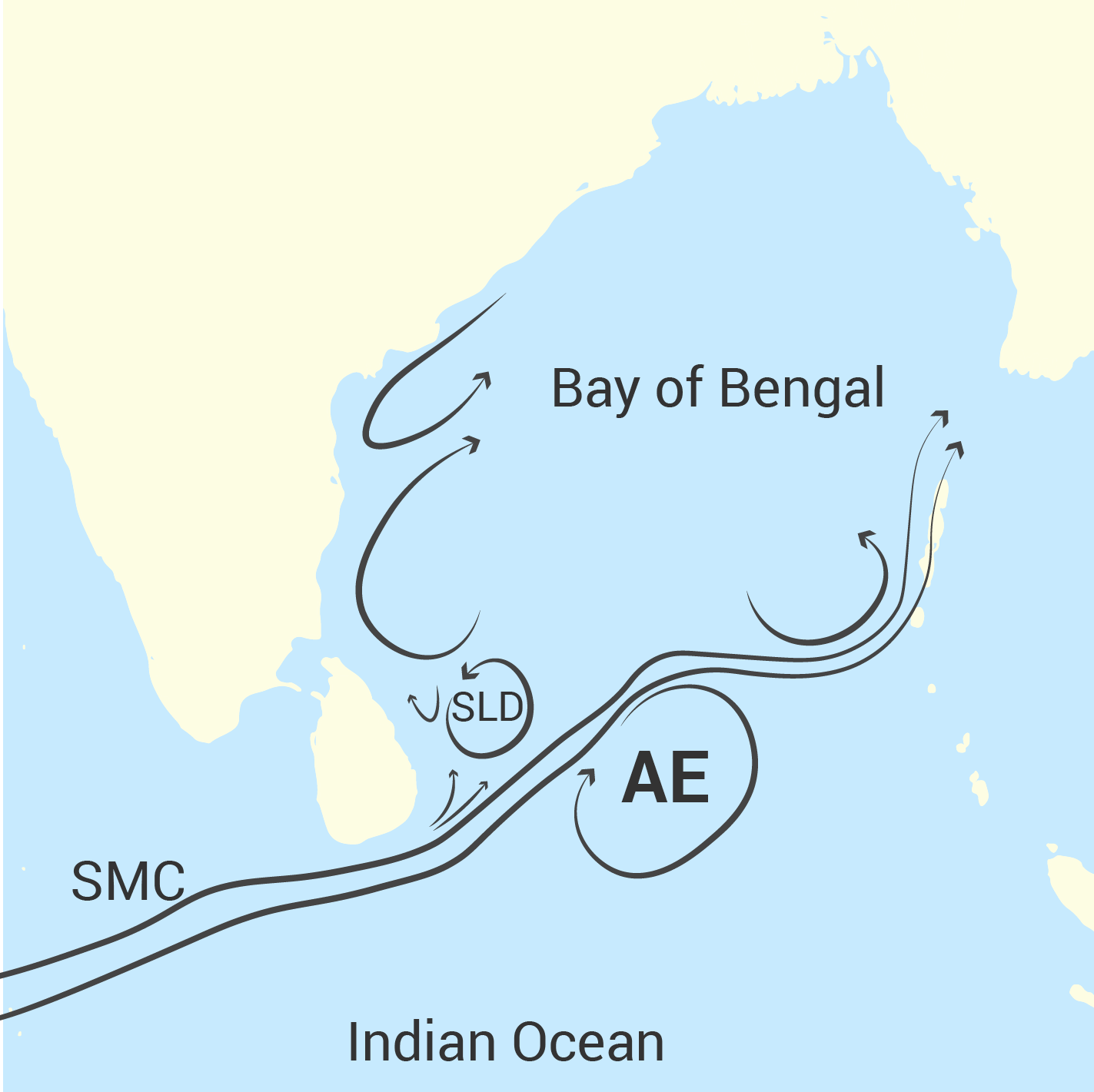}
    \caption{Currents and eddies in the BoB during the monsoon season, including the Summer Monsoon Current (SMC), the Sri Lanka Dome (SLD), and an anticyclonic eddy (AE).}
    \label{fig:bobSketch}
\end{figure}
The ocean circulation in the BoB is complex owing to the large amount of fresh water that enters the northern part of the bay and the seasonally reversing monsoon wind forcing. A river plume flows equator ward along the northern part of the east coast of India but the currents are oriented in the opposite direction in the southern part of the bay. Figure~\ref{fig:bobSketch} presents a rough schematic of the major currents and eddies in the bay during the monsoon season. The Summer Monsoon Current (SMC), a prominent feature of Indian ocean circulation, flows around Sri Lanka and into the BoB. In this section, we describe the use of pyParaOcean to study different phenomena in the BoB, particularly during the monsoon. This study demonstrates the utility of pyParaOcean in the study of ocean systems and is also of independent interest in terms of observations regarding the ocean structures in the BoB.

\begin{figure*}[!ht]
\centering
    \subcaptionbox{\label{subfig:cs3:1}}{\includegraphics[width=.33\linewidth]{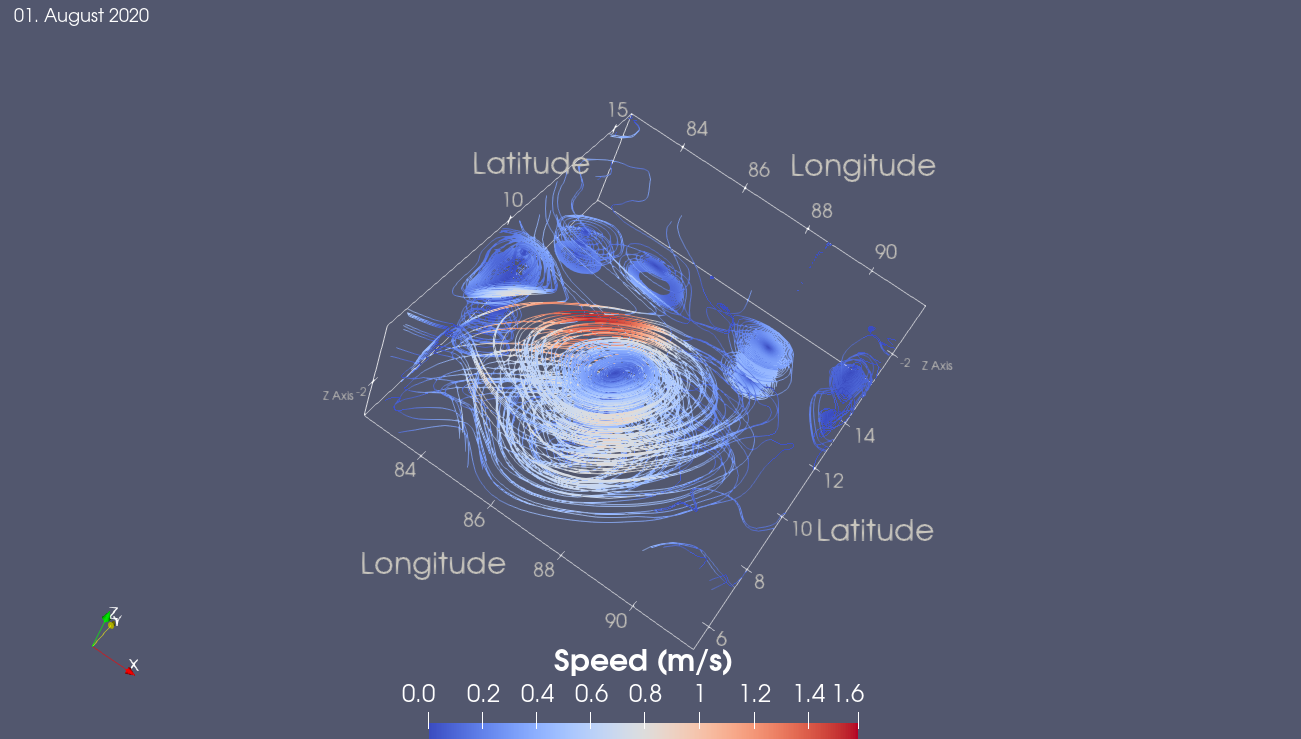}}
    \subcaptionbox{\label{subfig:cs3:2}}{\includegraphics[width=.33\linewidth]{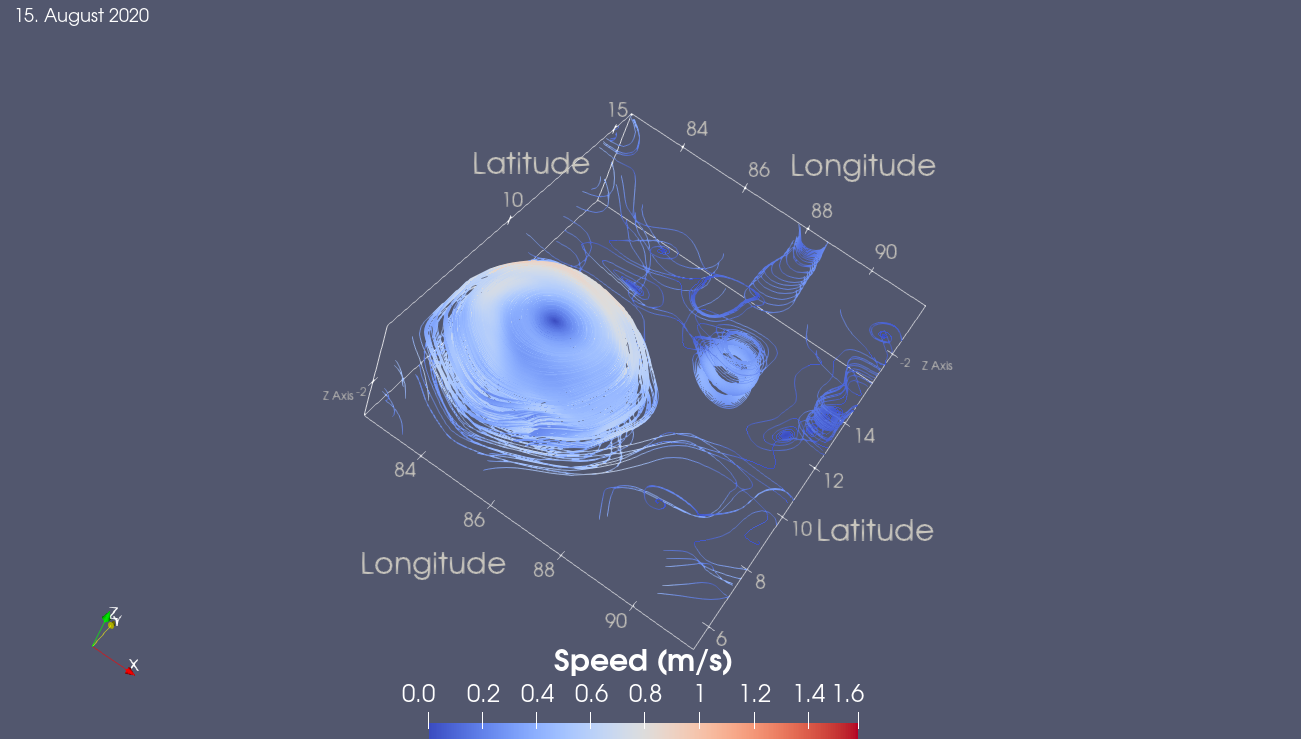}}
    \subcaptionbox{\label{subfig:cs3:3}}{\includegraphics[width=.33\linewidth]{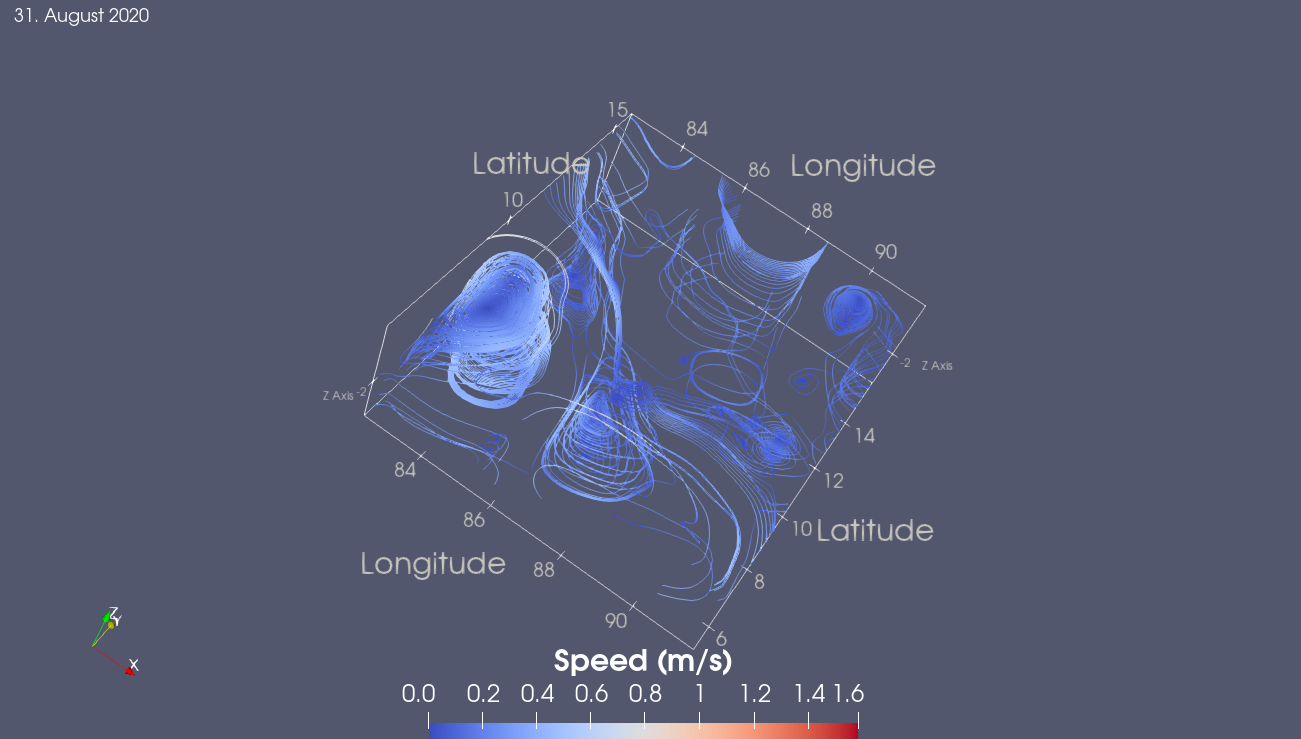}}
    \caption{Dissipation of a large anticyclonic eddy in the BoB during August 2020. Streamlines with seeds near detected vortex cores are computed to show the evolution of eddy profiles in 3D.}
   \label{fig:cs3}
\end{figure*}
\begin{figure*}[!ht]
\centering
    \subcaptionbox{\label{subfig:cs1:1}}{\includegraphics[width=.33\linewidth]{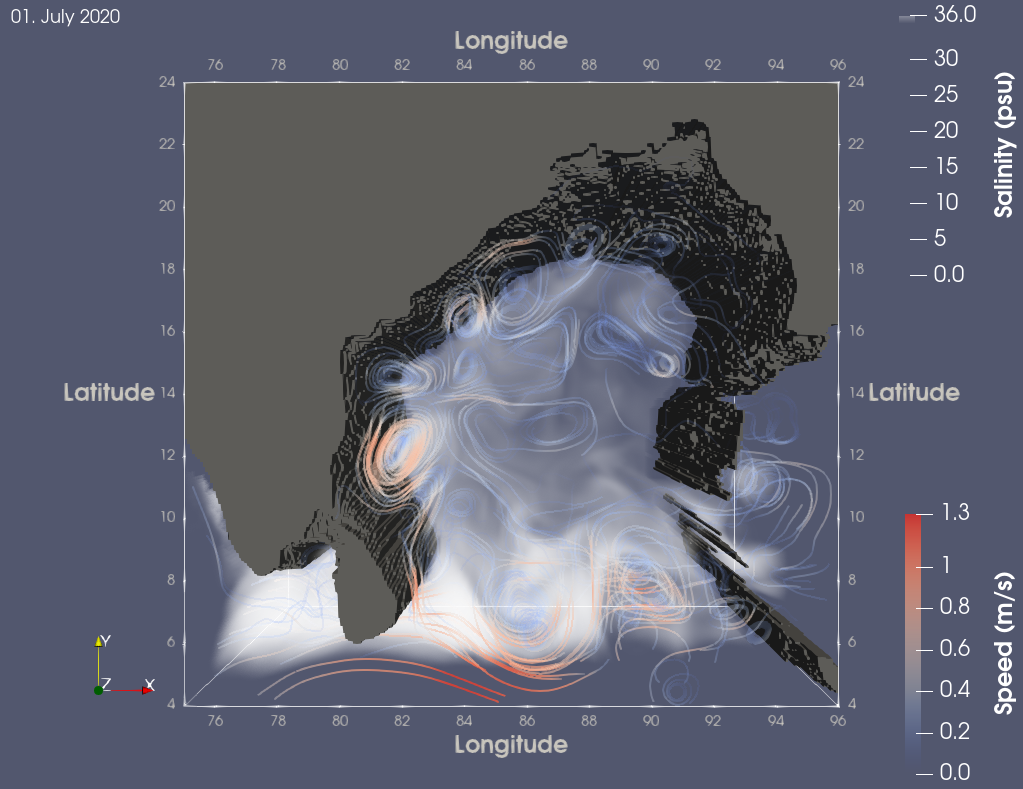}}
    \subcaptionbox{\label{subfig:cs1:2}}{\includegraphics[width=.33\linewidth]{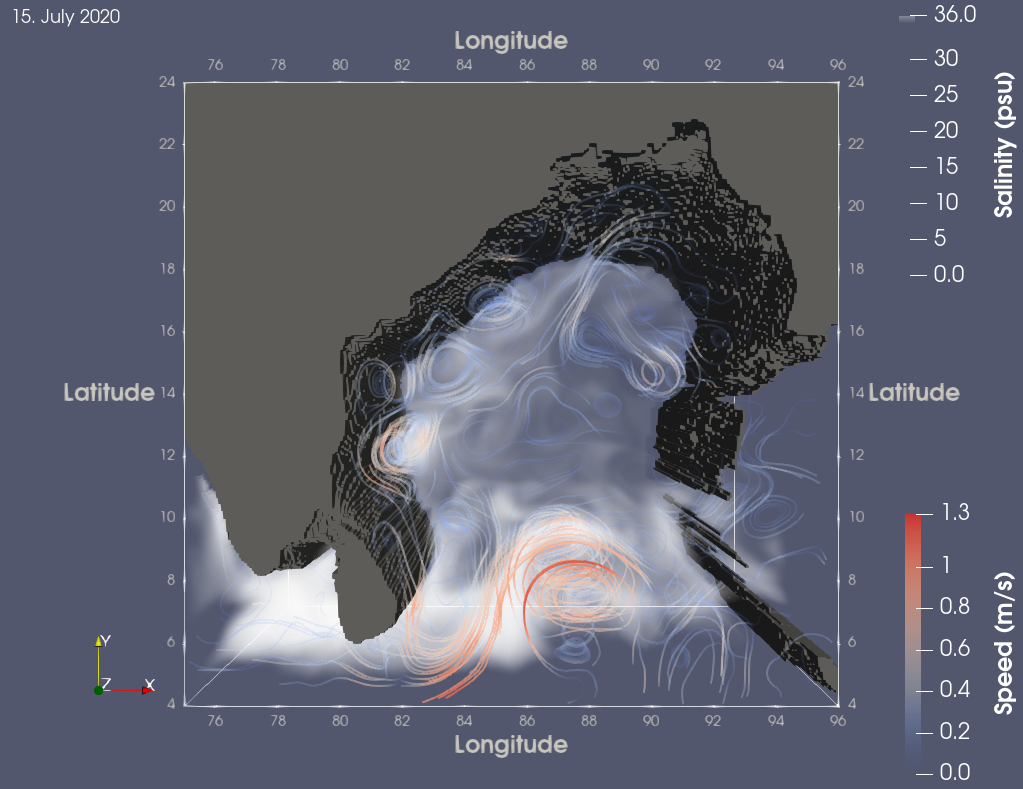}}    
    \subcaptionbox{\label{subfig:cs1:3}}{\includegraphics[width=.33\linewidth]{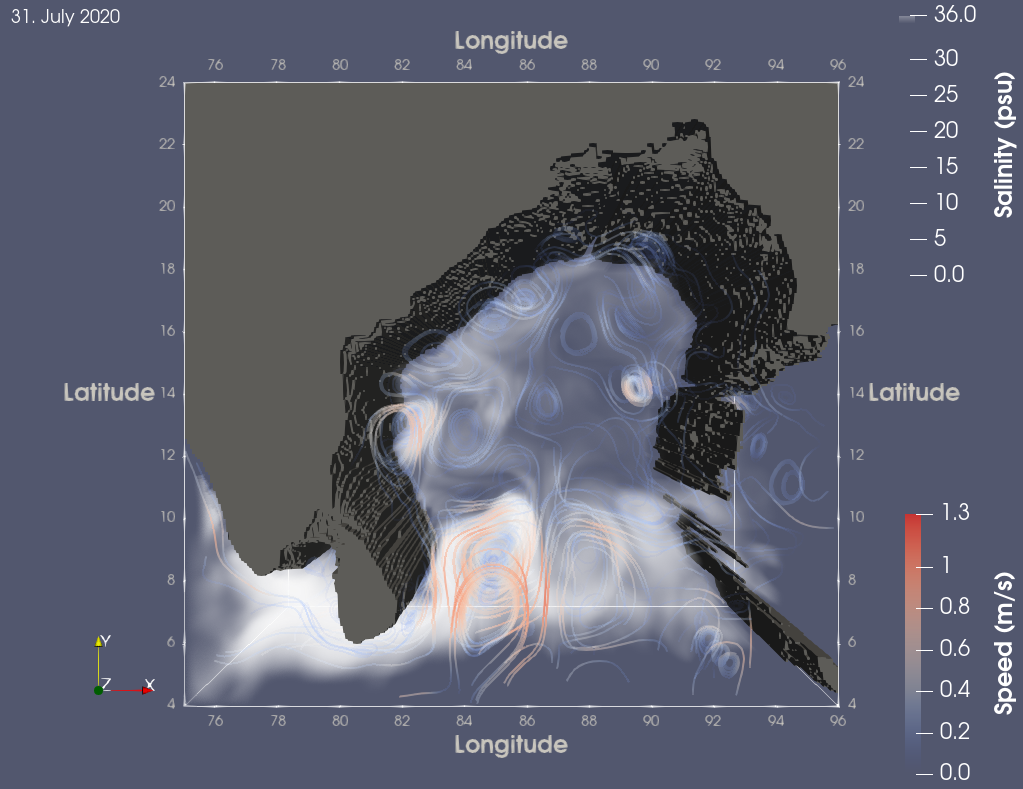}}
    \caption{The BoB between July 1, 2020 and July 31, 2020. Visualization of the flow using streamlines with uniform seeding and the $\geq 35$~psu salinity isovolume. (a)~July 1, 2020: The AE is forming around 8\textdegree N and 90\textdegree E with the SMC streamlines visible from 78\textdegree E to 86\textdegree E. (b)~July 15, 2020: The AE, 8\textdegree N and 87\textdegree E, has matured into a circular shape and moves westward towards Sri Lanka. The $\geq 35$~psu isovolume shows a recirculation of high salinity waters into the Bay by AE. (c)~July 31, 2020: The AE, 7\textdegree N and 84\textdegree E, reaches the eastern coast of Sri Lanka where it begins dissipating.}
    \label{fig:cs1}
\end{figure*}
\begin{figure*}[ht]
    \centering
    \includegraphics[width=.22\linewidth]{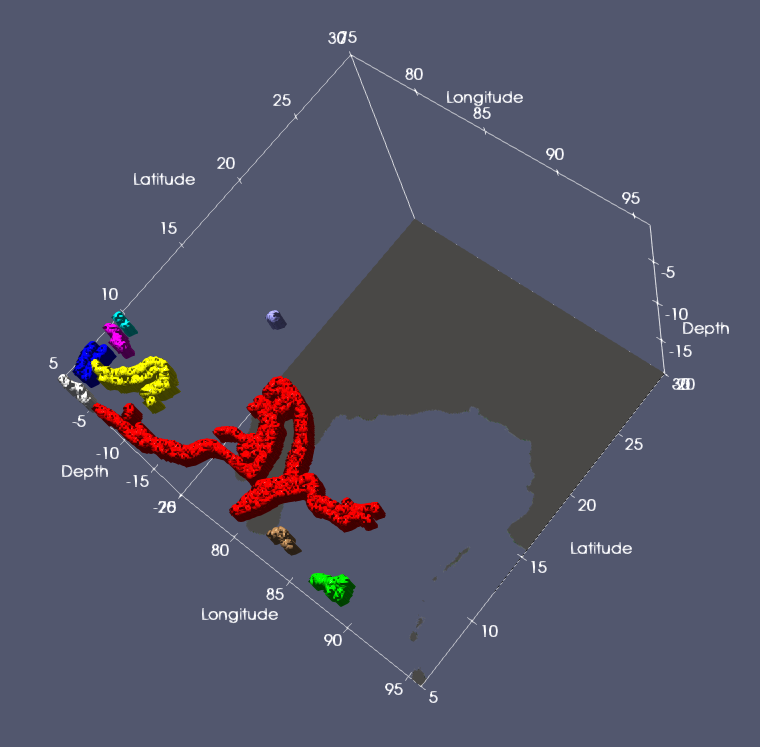}
    \includegraphics[width=.22\linewidth]{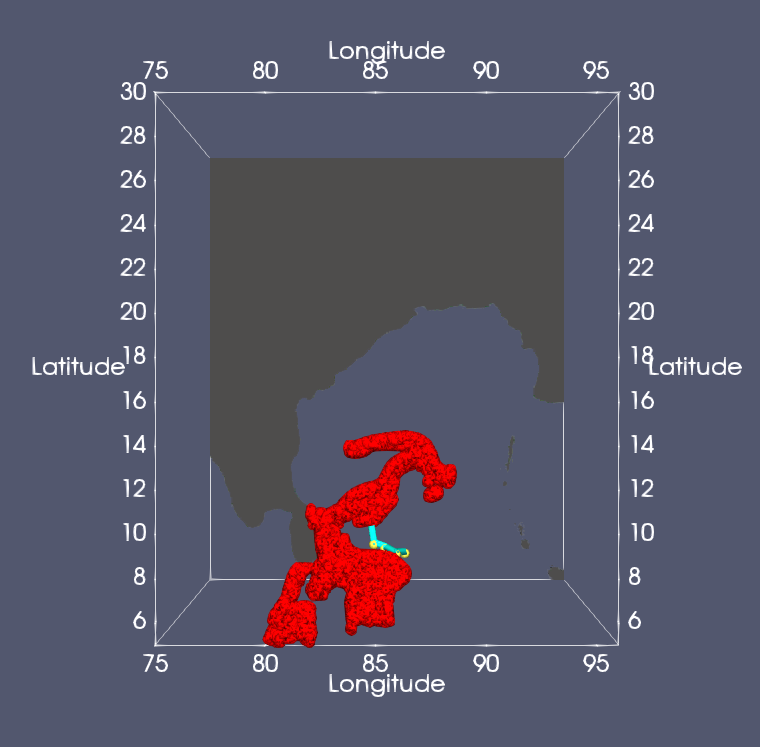}
    \includegraphics[width=.22\linewidth]{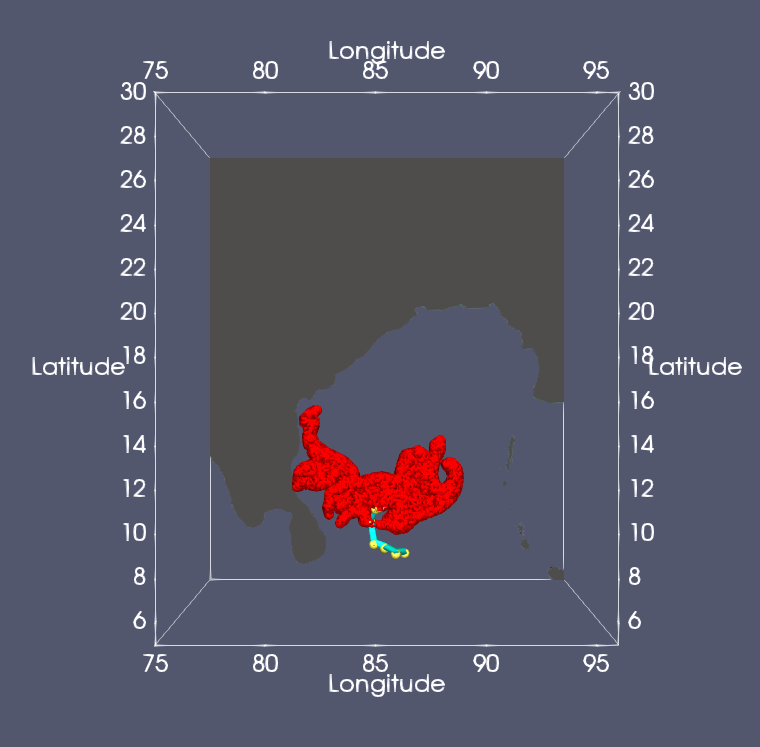}
    \includegraphics[width=.22\linewidth]{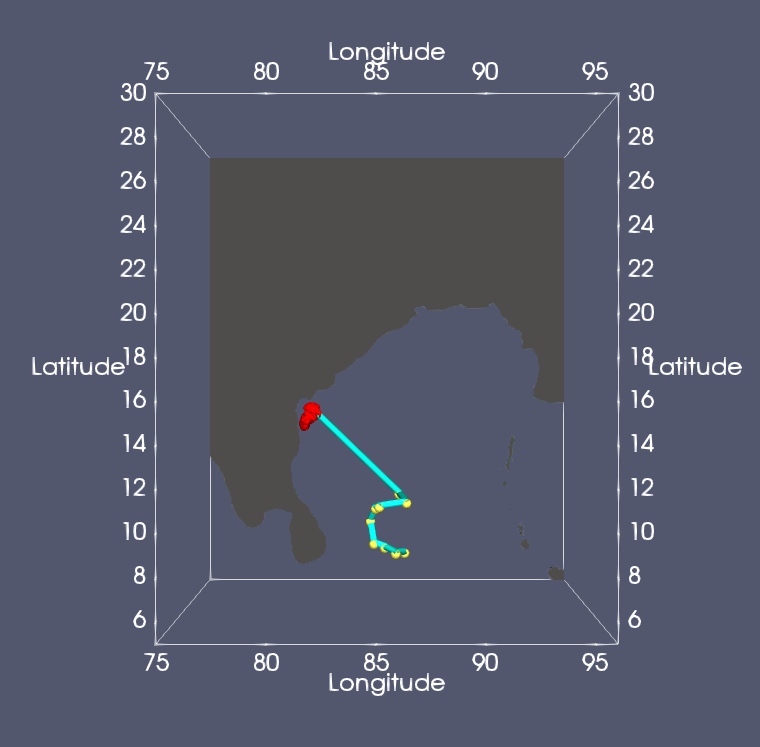}
    \caption{Visualizing movement of high salinity water via computation and tracking of surface fronts of high salinity isovolumes. (left)~Surface fronts computed at one time step. (middle, right)~One of the components of the surface front moves towards the east coast of India, near Visakhapatnam. The evolution of this surface front component is computed and visualized as a track.}
    \label{fig:hsc-front-track}
\end{figure*}
\begin{figure*}[!ht]
\centering
    \subcaptionbox{\label{subfig:cs2:1}}{\includegraphics[width=.45\linewidth]{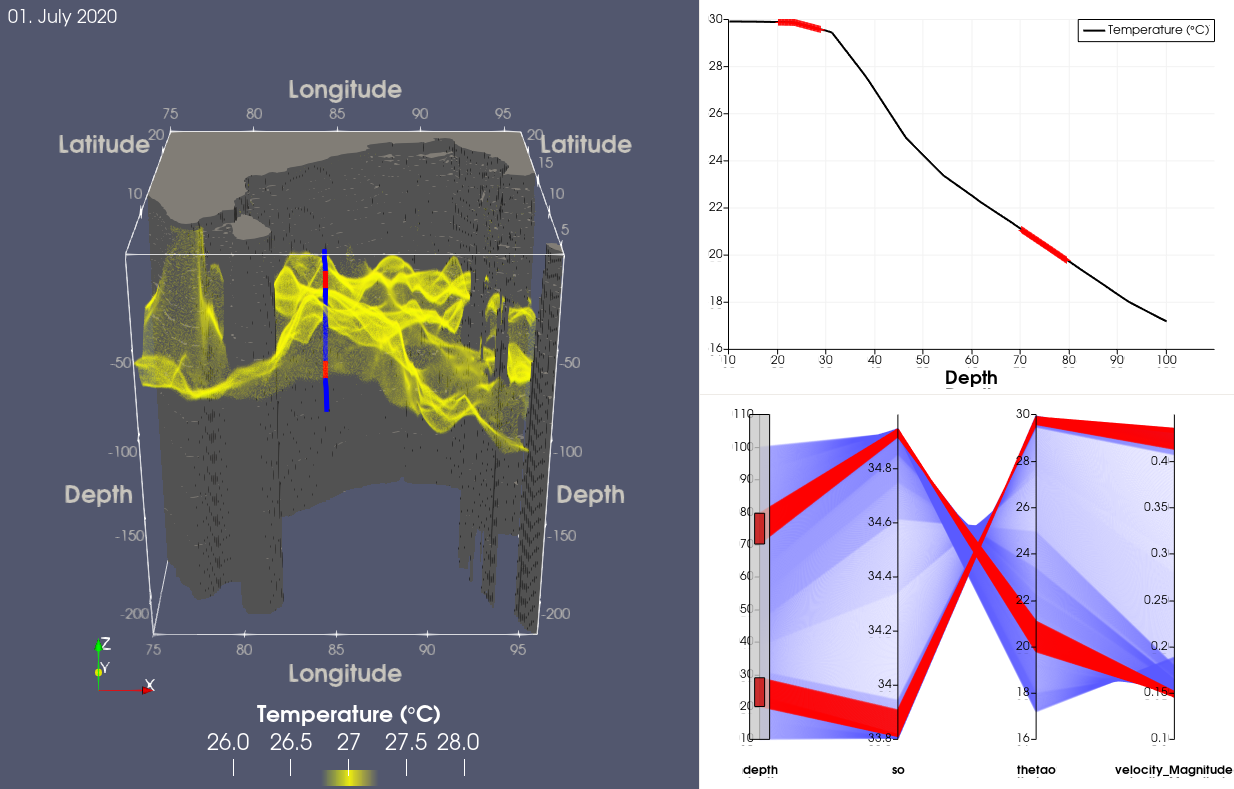}}
    \subcaptionbox{\label{subfig:cs2:2}}{\includegraphics[width=.45\linewidth]{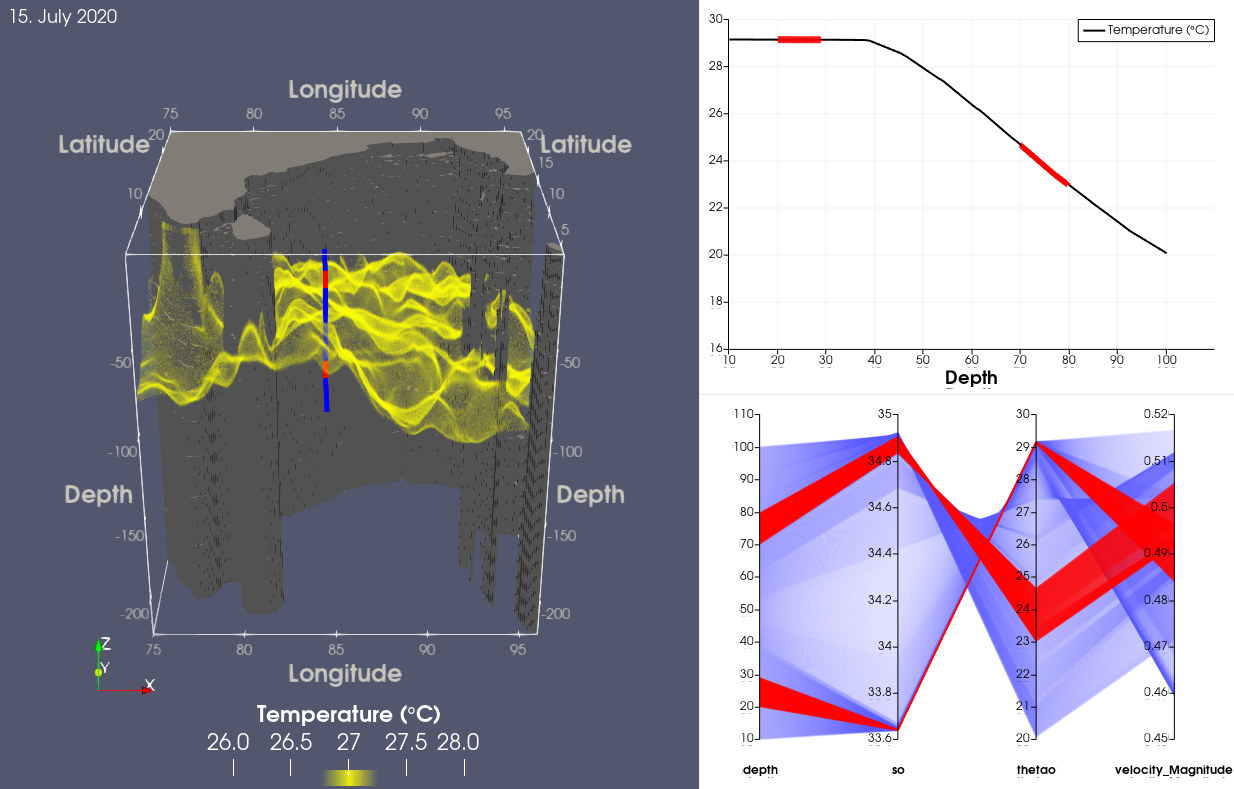}}
    \subcaptionbox{\label{subfig:cs2:3}}{\includegraphics[width=.45\linewidth]{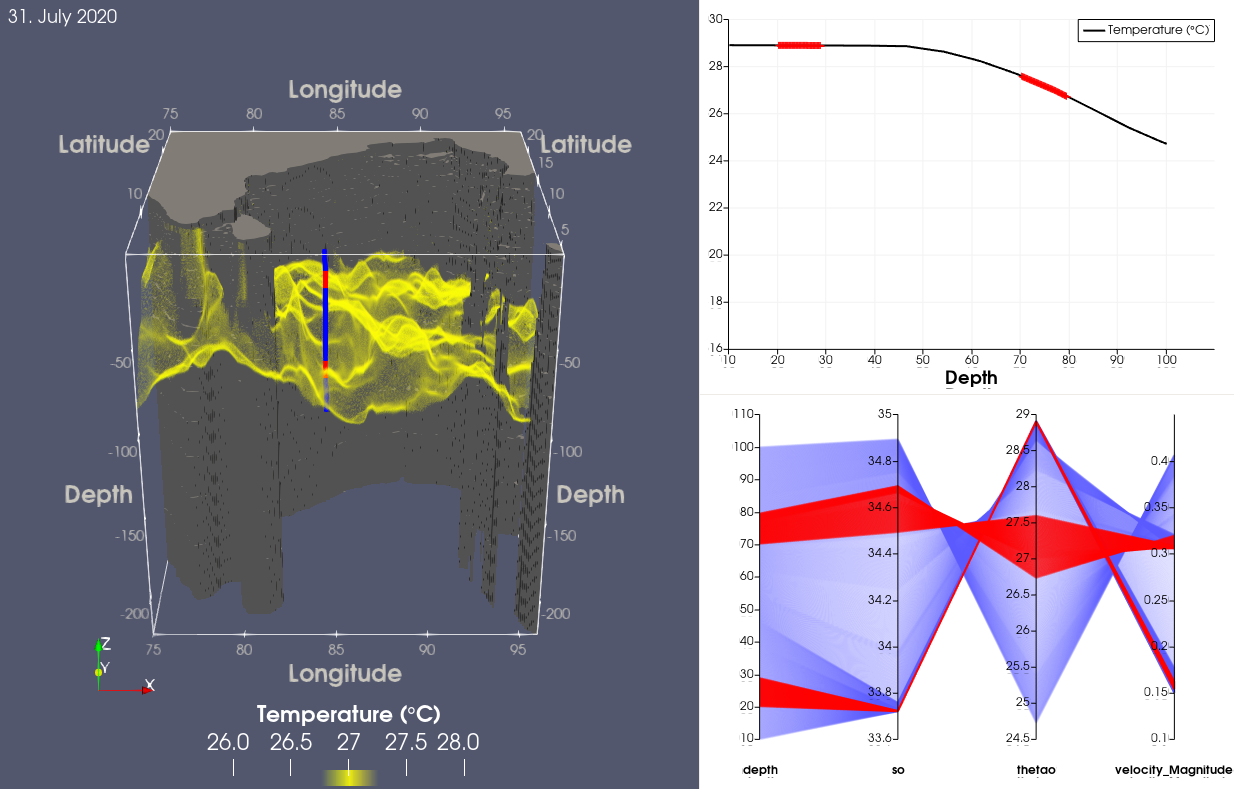}}
    \caption{The depression of the 27$^{\circ}$ C isotherm (yellow) by the anticyclonic eddy in the BoB. A needle is dropped at 7$^{\circ}$ N, 84$^{\circ}$ E and the depth profile shows the temperature drop. The interactive parallel coordinates plot is used to brush-select 10~m intervals at depths of 25~m and 85~m. (a)~July 1, 2020: The downwelling of the AE can be seen around 8$^{\circ}$ N and 90$^{\circ}$ E at the depth of 100~m. As it forms, the AE pushes the 27$^{\circ}$ isotherm down. (b)~July 15, 2020: The AE, 8$^{\circ}$ N and 87$^{\circ}$ E, can be seen moving east with the depression of the isotherm and the depth profile of temperature begins to flatten near 29$^{\circ}$C as the eddy moves closer to the needle. (c)~July 31, 2020: The AE center, 7$^{\circ}$ N and 85$^{\circ}$ E, is very close to the needle and the depression in the isotherm has moved all the way to near the east coast of Sri Lanka.}
    \label{fig:cs2}
\end{figure*}
\myparagraph{Eddies.} 
As shown in Figure~\ref{fig:bobSketch}, a large anticyclonic eddy (AE)  located to the right of the SMC and a cyclonic eddy known as the Sri Lanka Dome (SLD) to its left~\cite{vinayachandran1998monsoon} are regular features in this region during summer. The AE has a diameter of about 500~km, located to the southeast off the coast of Sri Lanka, and is characterized by intense downwelling inside owing to its anticyclonic circulation. Vinayachandran \etal~\cite{vinayachandran1998monsoon, rath2019dynamics} proposed that the AE is formed by the interaction of the SMC and the incoming Rossby waves from Sumatra. The timeline of the appearance and disappearance of the AE was documented in later work~\cite{vinayachandran2004biological}. We extract the eddy and visualize it using streamlines by applying the respective pyParaOcean filters with seeds placed near the vortex cores. The AE begins forming in June, develops into its circular shape in July, and weakens in August, as shown in Figure~\ref{fig:cs3} and the accompanying video. 

\myparagraph{Salinity transport.} 
The SMC carries high salinity water from the Arabian Sea into the BoB along its path. This supply of high salinity water is essential to maintain the salt balance of the bay. 
pyParaOcean serves as an efficient tool to analyze the effects of AE on the salt balance of the BoB. Streamlines and pathlines offer visualization of the circulation associated with the AE and its movement in the ocean. The fieldlines may be overlaid on a volume rendering of a scalar field to visualize the transport caused by the eddy. Figure~\ref{fig:cs1} and the accompanying video show the streamlines overlaid on a salinity volume rendering at different time steps to show the role of the AE in transport of salt. The movement of high salinity water from the Arabian sea by the SMC into the BoB and its recirculation by the AE is well captured in this representation. Tracking surface fronts of high salinity water and highlighting the long-lived tracks helps capture an overview of significant salinity movement in the region. We observe a track in Figure~\ref{fig:hsc-front-track} that moves towards the coast of India.

\myparagraph{Downwelling.} 
Figure~\ref{fig:cs2} and the accompanying video show the use of the depth profile filter to visualize the depression of the 27$^{\circ}$C isotherm by the AE. The anticyclonic nature of the eddy causes a downwelling inside the eddy and pushes the relatively warmer water downward. The parallel coordinates view shows changes in temperature, salinity, and speed in the water column caused by the arrival of the eddy at the point of interest.

\myparagraph{Filaments.}
\begin{figure*}
\centering
    \subcaptionbox{\label{subfig:filares:highres}}{\includegraphics[width=.49\linewidth]{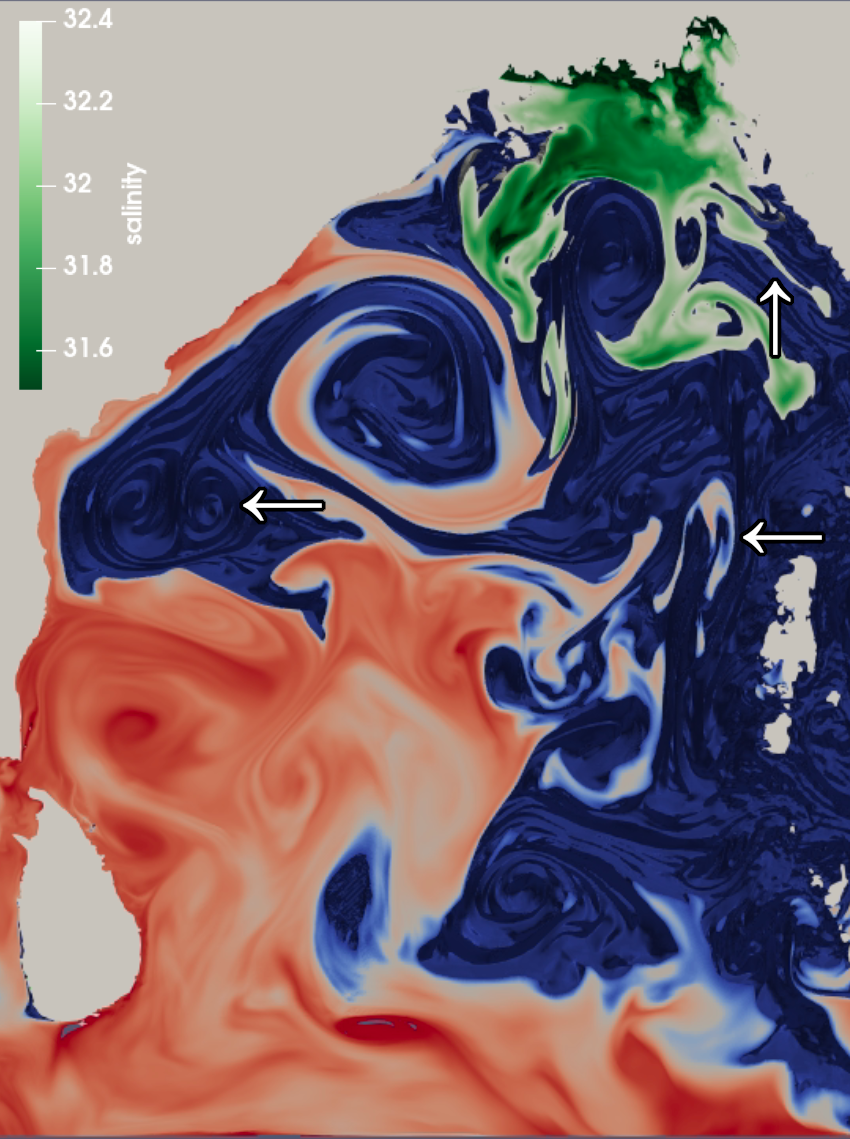}}
    \subcaptionbox{\label{subfig:filares:lowres}}{\includegraphics[width=.49\linewidth]{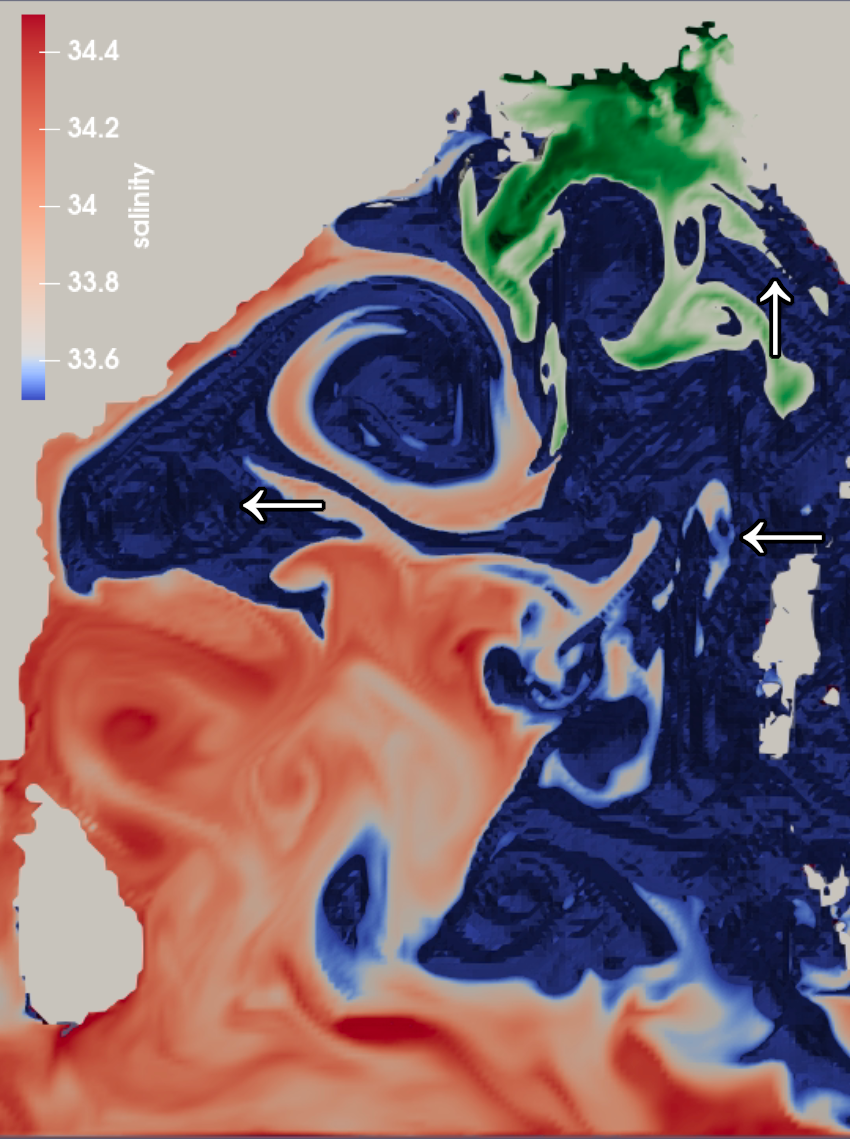}}
    \caption{The need for higher resolution data to extract certain ocean structures. (a)~A low salinity isovolume (green colormap) and a higher salinity isovolume (blue-red colormap) extracted from the ROMS data at a resolution of 96$^{th}$ of a degree clearly depicts filaments and some eddies (white arrows) in the BoB. (b)~Isovolume extracted from the data at a resolution of 12$^{th}$ of a degree. The filaments and eddies are broken up and the clarity of the structures are lost.}
   \label{fig:filares}
\end{figure*}
\begin{figure*}
\centering
    \subcaptionbox{\label{subfig:fila:1}}{\includegraphics[width=0.25\linewidth]{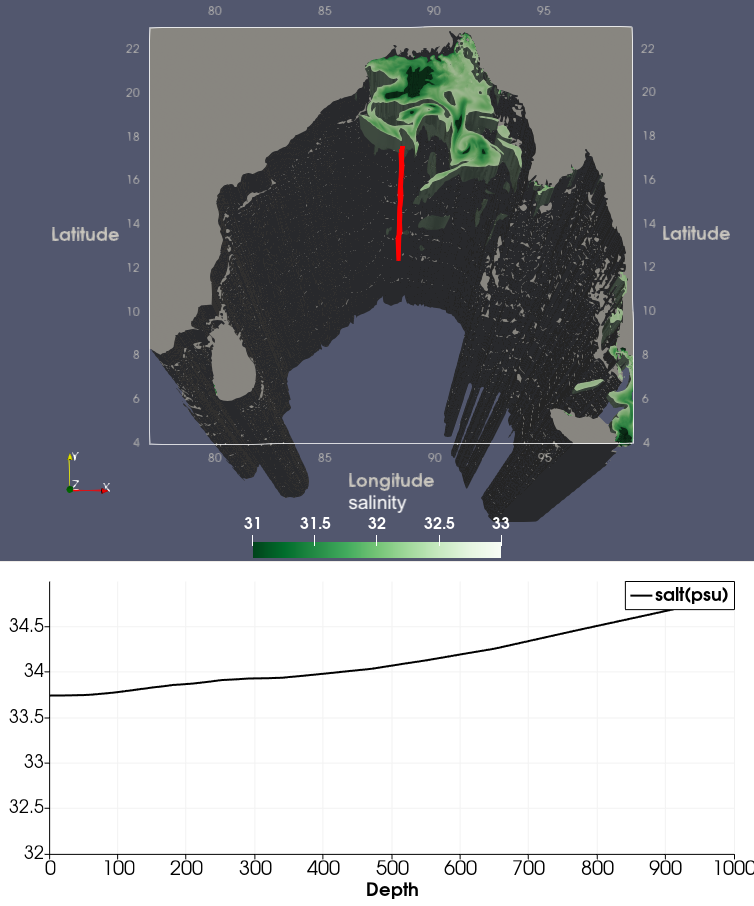}}
    \subcaptionbox{\label{subfig:fila:2}}{\includegraphics[width=0.25\linewidth]{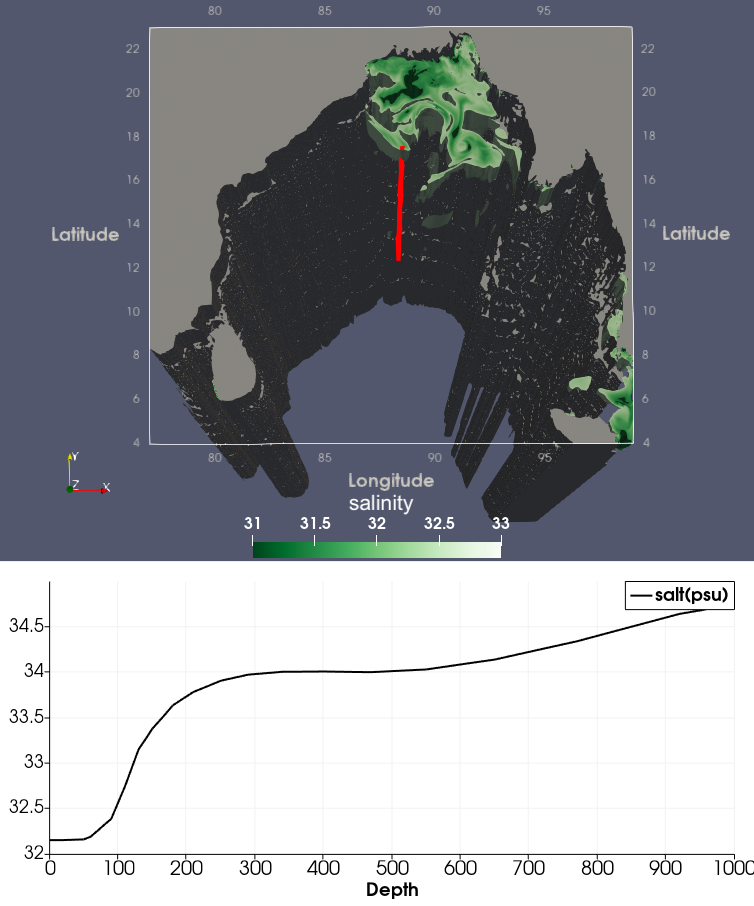}}
    \subcaptionbox{\label{subfig:fila:3}}{\includegraphics[width=0.216\linewidth]{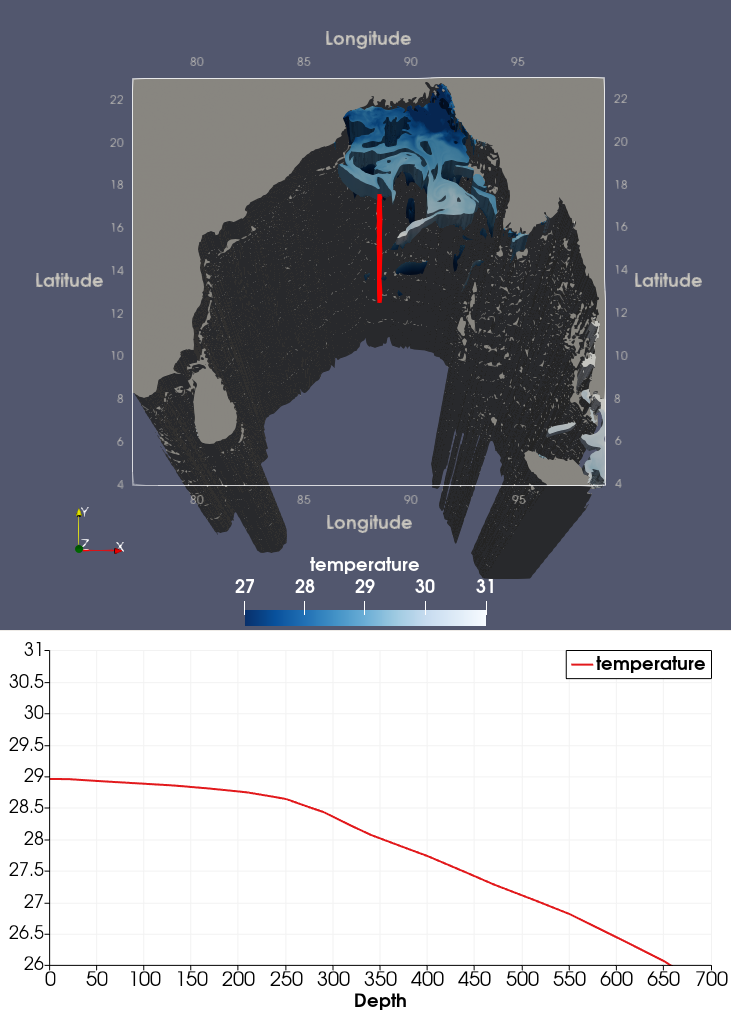}}
    \subcaptionbox{\label{subfig:fila:4}}{\includegraphics[width=0.216\linewidth]{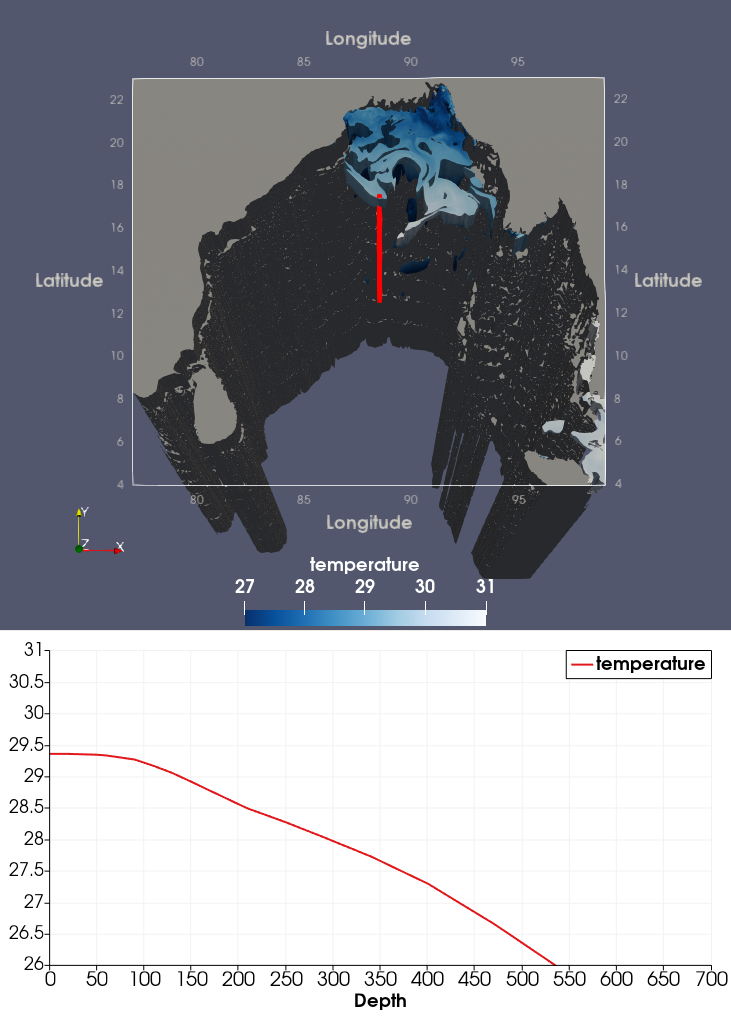}}
    \caption{\revision{Visualizing the temporal behavior of a low salinity filament using the depth profile filter on April~13 and~15, 2012.} (a,b)~Needle placed at 17.5$^{\circ}$ N, 88.5$^{\circ}$ E. Plot of salinity vs. depth and analysis of the filament isovolume may help in the study of barrier layers. (c,d)~Evolution of temperature distribution within the filament is studied using isovolume extraction, volume rendering, and the depth profile plot.} 
    
   \label{fig:fila}
\end{figure*}
An ocean filament is a long narrow strand of moving water within the ocean. Filaments can transport heat, nutrients, and marine life across vast distances. They are formed by wind patterns, currents, and differences in water density. Filaments act like rivers in the ocean, transporting water with unique characteristics, such as temperature, salinity, nutrients, and marine life, across long distances. These elongated features can stretch for hundreds of kilometers, while their width is typically only in the order of 10 kilometers and last from a few days to weeks. Due to their small scale, these features are difficult to analyze in a low resolution dataset. We demonstrate the need for a high resolution dataset such as ROMS in Figure~\ref{fig:filares}. Two isovolumes are rendered in Figure~\ref{fig:filares} - a low salinity~(31.5 - 32.5 psu) isovolume is represented with a green colormap and a higher salinity~(33.5 - 34.5 psu) isovolume is depicted by the blue-red colormap. Figure~\ref{subfig:filares:highres} shows the isovolumes in the ROMS dataset at a resolution of 96$^{th}$ of a degree, in which the filaments are legible. The same filaments are  not as clearly visible at a lower resolution of 12$^{th}$ of a degree of the same dataset in Figure~\ref{subfig:filares:lowres}. To demonstrate this further with examples, the three arrows point to \textbf{(a)}~a green colored low-salinity filament that is intact in Figure~\ref{subfig:filares:highres}, but broken up in Figure~\ref{subfig:filares:lowres}, \textbf{(b)}~a red colored isovolume that loses its shape in the lower resolution dataset, and, \textbf{(c)}~two eddies in dark blue color that lose their structure in the lower resolution dataset. Such filaments and fine features belong to a class of so-called submesoscale features in the ocean. The tools available in pyParaOcean support the efficient tracing of  filaments and other submesoscale structures and hence facilitate further statistical study to determine their impact on the ocean environment.

We use the depth profile filter to analyze the behavior of a low salinity filament in the northern BoB. We drop a needle at 17.5$^{\circ}$ N, 88.5$^{\circ}$ E and analyze the behavior of salinity using an Eulerian method. This involves the study of the salinity behavior on the needle across different time steps. We also display an isovolume that contains the low salinity filament, colored based on the salinity distribution. Figure~\ref{subfig:fila:1} shows the behavior of salinity on April 13, 2012 in the salinity-depth plot. The salinity increases steadily at this time step as we go deeper into the ocean. This is due to the higher salinity water having more density. In the next time step at April 15, shown in Figure \ref{subfig:fila:2}, we observe that the salinity isovolume crosses the needle, resulting in a sudden drop in salinity in the depth plot. The filament affects the salinity of the water only within the top 200~m and we see the normal steady increase in salinity as we go deeper. The local drop in salinity can lead to the formation of barrier layers~\cite{vinayachandran2002observations} and the shallowing of mixed layers affecting the air-sea interaction. 

Next, we study the temperature distribution within this filament as it crosses the same needle. Figures~\ref{subfig:fila:3} and~\ref{subfig:fila:4} correspond to the April~13 and April~15 time steps, respectively. The temperature distribution is visualized using a volume rendering of the salinity isovolume and a depth plot. We observe from the temperature-depth plot that the temperature rises by 0.5$^{\circ}$~C when the isovolume crosses the needle. 
The extraction of the filament using one property (salinity) followed by its visualization in terms of the time evolution of the temperature field is naturally supported by pyParaOcean filters. This analysis throws some light on how the filament interacts with the surroundings. In this case, the salinity of the filament does not change as it moves, which implies that the mixing of the filament with the surrounding water is low. But, we observe from the volume rendering that the filament does not hold its temperature. The change in temperature over time can be attributed to the interaction of the filament with the atmosphere. This method can be extended to determine exchanges between coastal and open ocean waters, and can be helpful in tracking tracers like sediments and chlorophyll in the fresher water filaments or for tracking oil spills and microplastics to identify the impact of these tracers on the ocean. Quantification of the magnitude and spatial extent of such changes in physical parameters can impact predictions by ocean and atmospheric models.

\revision{\section{Discussion}
The scalability experiments and case study validate the utility and applicability of pyParaOcean to large datasets, with potential for broader use in geoscience applications. We now present user inputs and discuss extendibility of the system with respect to the system design and visualization methodology.  

\myparagraph{User experience.}
This case study was conducted in collaboration with a senior oceanographer coauthor. Below are some comments from them regarding the significance of the study and the advantages of using pyParaOcean for the visual analysis tasks -- ``Analyzing high-resolution ocean model outputs is a challenging task. In addition to the large volume of the time-dependent 3D dataset, the analysis task is further complicated when multiple variables have to be visualized and associations amongst them need to be examined. The combination of Cinema Viewer and the isovolume and depth profile representations provide a relatively handy approach to plow through the large volume of data. As an example, the inflow of large volume of fresh water from river systems such as Amazon and Ganga-Brahmaputra are well known. These rivers also bring along other substances such as sediments and nutrients. Understanding the distribution of these foreign matter is crucial for monitoring the health of the oceans and their impact on the marine system. In the case study, we have shown how temperature and salinity associated with such inputs are tracked and a preliminary assessment of their interaction with the surroundings can be made. A similar approach can be adopted for other passive and active tracers in the ocean and to assess the interaction between different species of tracers. The tools developed here can be used to investigate upwelling fronts or the transport along the periphery of eddies. The Cinema database is detailed enough to get meaningful insights from the data and compact enough to be carried in a USB flash drive.''

Anecdotal inputs from two oceanographers, including one coauthor, suggest that the installation of pyParaOcean is straightforward, and a mild learning curve is sufficient to understand and get acquainted with the visualization pipeline of ParaView. Notably, the interactivity of the system is a significant advantage. While our oceanographer collaborators typically use tools such as pyFerret for 2D analysis, they found the capabilities of pyParaOcean to be very useful and easy to use. In their typical workflow, identifying or studying a phenomenon involved analyzing individual time steps in a NetCDF viewer before importing them into Ferret. The Cinema viewer was an efficient alternative, offering a portable and lightweight solution for quick tasks while additionally supporting the detailed and interactive study of the time evolution of 3D fields. Beyond the functionalities provided by pyParaOcean, the built-in features of ParaView, such as extracting a subset, transfer function editing, calculator tools, and slicing the domain of a dataset proved to be incredibly useful in their analysis workflow.

\myparagraph{System Design.}
We developed pyParaOcean as a plugin for the open source software ParaView, to capitalize on its robust scalability support through OpenMPI and advanced visualization capabilities. As demonstrated in the previous sections, this choice results in strong scaling performance and provides a flexible foundation for adaptation across various geoscience applications. While ParaView offers an extensive array of visualization techniques, it is not entirely self-sufficient; additional tools are necessary to bridge certain functionality gaps. For instance, we incorporate the Cinema Science generator as an auxiliary tool, enabling users to analyze the data and tasks at hand before launching the resource-intensive filters within ParaView.

Our implementation of pyParaOcean filters relies, to a large extent, on existing VTK implementations with the objective of computational efficiency and conformation to latest standards. For more complex computation, such as the eddy identification and visualization filter, we designed the filters such that they support multiple implementations catering to the data, the available scalar and vector fields and the task at hand; see subsection~\ref{subsec:eddyidentificationtool}.

Another key factor for selecting ParaView is its server-client architecture, which provides flexibility to independently deploy the client, the render server, and the data server. This modular architecture enables users to reliably scale to large datasets and efficiently distribute the compute resources at their disposal. Importantly, the workflow, visualization pipeline, and interface presented to the user remain consistent regardless of deployment configuration and thus ensures seamless user experience.

\myparagraph{Visualization Design.}
The design of the visualization modules in pyParaOcean was guided by continuous feedback from our oceanographer coauthor to ensure they meet practical needs in ocean data analysis. The filters described above address many of the core tasks essential for exploring and analyzing ocean datasets. We envision pyParaOcean as an evolving system that will support additional tasks in future, possibly for other geoscience applications. Central to the plugin is its visualization pipeline, inherited from ParaView, which provides users with an intuitive GUI to layer multiple computational steps. The pipeline design abstracts much of the complexity behind these tasks, while also giving advanced users the option to engage with more sophisticated analyses using tools like the programmable filter and the Python shell. This extensibility enables contributions that further expand the capabilities of the system, such as from users towards oceanography or other geoscience applications. The seemingly trivial routine tasks such as data slicing, calculating derived scalars and vectors, and 3D rendering are already built into ParaView and can be easily included into the pipeline. Our visualization choices were further shaped by a commitment to maintaining interactivity even when the system is deployed across multiple cores in a cluster environment. pyParaOcean aims to keep users in control by continuously displaying computation output and enabling intuitive, seamless interaction with data throughout the analysis workflow.}

\section{Conclusions}
This paper presented an interactive and scalable system, called pyParaOcean, for the visualization of 3D time-varying fields in oceanography. A detailed case study helped confirm the utility of pyParaOcean and a comprehensive scaling study demonstrated its applicability of its modules to large data sizes. 
In future, we plan to incorporate fast and efficient parallel implementations of algorithms for computing other ocean structures of interest and for tracking them across time. Identification and tracking of features represented by scalar fields is a recurring problem in different subfields of geoscience. The design approach and methods described in this paper may be extended to study datasets from related application domains such as atmospheric science and meteorology.
pyParaOcean is available for download in the public domain for use by the community~\cite{pyParaOcean}.

\section*{Acknowledgments}
This research was funded by a grant from SERB, Govt. of India (CRG/2021/005278), partial support to PNV from National Supercomputing Mission, DST, Govt. of India, the Dr. Ram Kumar IISc Distinguished Visiting Chair Professorship in EECS, and a scholarship from MoE, Govt. of India. VN acknowledges support from the Alexander von Humboldt Foundation, and Berlin MATH+ under the Visiting Scholar program. Part of this work was completed when VN was a guest Professor at the Zuse Institute Berlin.

\noindent \textbf{Data Availability.} The data that support the findings of this study are available from the corresponding author upon reasonable request.

\noindent \textbf{Conflict of Interest.} None.

\bibliographystyle{eg-alpha-doi} 
\bibliography{scalable}       


\end{document}